\documentclass[preprint,12pt]{article}
\usepackage{amsmath, amssymb}
\usepackage{graphics}
\def\mylimit#1{\mathrel{\mathop{\kern0pt\longrightarrow}\limits_{#1}}}

\newcommand{\VEV}[1]{\left\langle #1 \right\rangle}

\newcommand{\nn}{\nonumber}

\newcommand{\order}[1]{{\cal O}(#1)}

\newcommand{\GeV}{\mbox{GeV}}

\newcommand{\abs}[1]{\left| #1 \right|}
\newcommand{\s}[1]{\widetilde{#1}}

\newcommand{\cc}[1]{\overline{#1}}

\newcommand{\eff}{{\mrm {eff}}}

\newcommand{\bequ}{\begin{equation}}
\newcommand{\eequ}{\end{equation}}
\newcommand{\beqn}{\begin{eqnarray}}
\newcommand{\eeqn}{\end{eqnarray}}
\newcommand{\bctr}{\begin{center}}
\newcommand{\ectr}{\end{center}}
\newcommand{\bit}{\begin{itemize}}
\newcommand{\eit}{\end{itemize}}
\newcommand{\Ls}{\left(}
\newcommand{\Rs}{\right)}
\newcommand{\Lm}{\left\{}

\newcommand{\Ll}{\left[}
\newcommand{\Rl}{\right]}

\newcommand{\RR}{\right.}
\newcommand{\hsp}[1]{\hspace {#1cm}}

\newcommand{\half}{{1\over2}}
\newcommand{\mrm}{\rm}

\begin{document}
\begin{titlepage}

\begin{flushright}
hep-ph/0404020\\
KUNS-1906\\
\today
\end{flushright}

\vspace{4ex}

\begin{center}
{\large \bf
Horizontal symmetry in Higgs sector of GUT with $U(1)_A$ symmetry
}

\vspace{6ex}

\renewcommand{\thefootnote}{\alph{footnote}}

Nobuhiro Maekawa\footnote
{E-mail: maekawa@gauge.scphys.kyoto-u.ac.jp
}
and 
Toshifumi Yamashita\footnote
{E-mail: yamasita@gauge.scphys.kyoto-u.ac.jp}

\vspace{4ex}
{\it Department of Physics, Kyoto University,\\
     Kyoto 606-8502, Japan}\\
\end{center}

\renewcommand{\thefootnote}{\arabic{footnote}}
\setcounter{footnote}{0}
\vspace{6ex}

\begin{abstract}
In a series of papers, we pointed out that an anomalous $U(1)_A$ gauge 
symmetry naturally solves various problems in grand unified theories (GUTs)
and that a horizontal gauge symmetry, $SU(2)_H$ or $SU(3)_H$, not only realizes 
the unification of three generation quarks and leptons in fewer multiplets but
also solves the supersymmetric flavor problem. 
In this paper, we examine the possibility that the Higgs sectors of the 
GUT symmetry and of the horizontal symmetry  are unified,
that is, there are some Higgs fields whose vacuum expectation values (VEVs)
break
both the GUT gauge symmetry and the horizontal symmetry at the same time.
Although the scale of the VEVs 
become too large to suppress the flavor changing neutral current 
processes sufficiently, the unification is possible. In addition, 
for the $SU(3)_H$
models, the $SU(3)_H$ gauge anomaly is cancelled in the unified models
without introducing additional fields in contrast with the previous models
in which the Higgs sectors are not unified.
\end{abstract}

\end{titlepage}


\section{Introduction}
We have several reasons for introducing a horizontal symmetry $G_H$.
One of them is to understand the origin of the flavor violation
in Yukawa couplings of quarks and leptons. Actually many studies have
been done along this direction\cite{discrete}-\cite{nonabel}.

The second reason is to unify quarks and leptons in fewer multiplets,
though it is strongly related with the first motivation.
Usual Grand unified theory (GUT) has two kinds of unification. 
The first unification is to unify gauge interactions, and the 
second is to unify quarks and leptons. However, usual GUT with $SU(5)$,
$SO(10)$ or $E_6$ gauge group can unify only one generation quarks and
leptons. 
In order to unify all the quarks and leptons into a single multiplet,
a larger gauge group such as $SO(12+2n)$, $E_7$, or $E_8$ is required,
though these unified group cannot realize chiral matter 
in a 4-dimensional theory. 
However, actually it is possible in higher dimensional field
theories, and in that cases, a horizontal symmetry can appear in
the effective 4-dimensional field theories.

The third reason is to solve the flavor problem 
in supersymmetric theories\cite{FCNC}.
A non-abelian flavor (horizontal) symmetry can potentially 
 solve the problem. 
If the first
two generation fields become  a doublet under the flavor symmetry, 
$\Psi_a$($a=1,2$), it is obvious that, unless the flavor symmetry is broken,
the sfermion masses of the first two 
generation fields become
universal, which is important in solving the SUSY flavor problem. 
Of course, in
order to obtain realistic hierarchical structure of Yukawa couplings, 
the flavor symmetry must be broken, for example, 
 by the vacuum expectation value (VEV) $\VEV{F_a}$. 
Then, generically the universal sfermion
masses are lifted by the breaking. 
Various models in which the breaking
effects can be controlled have been considered 
in the literature\cite{Sdiscrete,Sabel,Snonabel}.
However, in GUT models with bi-large neutrino mixings, 
which have been reported by recent experiments\cite{atmos,solar}, 
the universality of the first two generation sfermion
masses is not sufficient
to solve the SUSY flavor problem, because the large 
mixings and the $\order1$ discrepancy between the sfermion masses 
of the third 
generation fields and those of the first two generation fields lead to 
too large flavor changing neutral current (FCNC) processes, for example, 
$\epsilon_K$ in $K$ meson, $\mu\rightarrow e\gamma$, 
$\tau\rightarrow \mu\gamma$, etc. One of the authors pointed out
that $E_6$ unification can naturally solve this problem 
in Refs.\cite{horiz}.
The essential point is that 
in the $E_6$ unification all three light modes of ${\bf \bar 5}$
fields come from the first two generation fields 
$\Psi_a({\bf 27})$\cite{Etwist}, 
because it is naturally expected 
that the ${\bf \bar 5}$ fields from the third 
generation field $\Psi_3$ become superheavy owing to its large Yukawa
couplings. Therefore, all the three light modes of ${\bf \bar 5}$ 
have universal sfermion masses, which are important 
in solving the SUSY flavor problem.

In a series of papers\cite{maekawa,BM,E6}\cite{maekawa2}-\cite{MY}, 
we proposed an attractive GUT scenario with an
anomalous $U(1)_A$ gauge symmetry\cite{U1}, 
whose anomaly is cancelled by the Green-Schwarz mechanism\cite{GS}. 
One of the biggest differences between our works and previous
works is to introduce generic
interactions even for higher dimensional interactions. Therefore,
once we fix the symmetry of the theory, 
we can define the theory except $\order1$ coefficients. 
It is interesting that under this natural assumption, 
the doublet-triplet (DT) splitting is realized 
with sufficiently stable nucleon, and 
realistic quark and lepton mass matrices including 
the bi-large neutrino mixings are obtained. 
Moreover, the non-abelian horizontal gauge symmetry can be 
naturally introduced in the GUT scenario
with an anomalous $U(1)_A$ gauge symmetry to solve 
the SUSY flavor problem,
especially in the $E_6$ unification. 

In Refs.\cite{horiz}, the Higgs fields to break the 
horizontal gauge symmetry
$G_H=SU(2)_H$ or $SU(3)_H$ are different 
from the Higgs fields to break
the GUT gauge group $SO(10)$ or $E_6$. 
However, in principle, these Higgs
sector can be unified, namely, a non-vanishing VEV
of a Higgs field  with non-trivial representation 
for both gauge groups can 
break both gauge groups. 
If it is realized, the number of the Higgs fields
can be reduced. 
In this paper, we examine this possibility.

After this introduction, we give a brief review of 
the horizontal symmetry
that is introduced to suppress the FCNC processes 
in section 2. 
And in section 3, we examine the possibility 
for $SO(10)$ unification,
and in section 4, $E_6$ unification is examined.

\section{Horizontal symmetry for SUSY flavor problem}
\label{horizreview}
In this section, we briefly review an idea that 
a horizontal symmetry is
introduced to solve the SUSY flavor problem, 
because this is one of the 
most important motivation to introduce a horizontal symmetry.

For simplicity,  we consider a simple model with 
a global horizontal symmetry $U(2)$, under which the three
generations of quarks and leptons,
$\Psi_i=(\Psi_a,\Psi_3)$ ($a=1,2$), are transformed as ${\bf 2+1}$, and
the Higgs field $H$ is a singlet.
Then only the Yukawa interaction $\Psi_3\Psi_3H$ is allowed 
by the horizontal 
symmetry, which accounts for the large top Yukawa coupling. 
Suppose that the $U(2)$ horizontal symmetry is broken by the 
VEVs of a doublet $\VEV{\bar F^a}=\delta^a_2V$ and of an anti-symmetric
tensor $\VEV{A^{ab}}=\epsilon^{ab}v$ ($\epsilon^{12}=-\epsilon^{21}=1$) as
\begin{equation}
  U(2)_H \mylimit{V}U(1)_H \mylimit{v} {\rm nothing}.
\end{equation}
The ratios of the VEVs to the cutoff, $\epsilon\equiv V/\Lambda\gg
\epsilon'\equiv v/\Lambda$, give the following hierarchical structure of
the Yukawa couplings:
\begin{equation}
Y_{u,d,e}\sim\left( 
 \begin{matrix}
  0 & \epsilon' & 0 \\
  \epsilon' & \epsilon^2 & \epsilon \\
  0 & \epsilon & 1
 \end{matrix}
\right).
\end{equation}
Moreover,
the $U(2)_H$ symmetric interaction
 $\int d^4\theta \Psi^{\dagger a}\Psi_a Z^\dagger Z$, where $Z$ 
 has a non-vanishing VEV given by $\VEV{Z}\sim \theta^2\tilde m$,
leads to approximate universality of the first and second generation 
sfermion masses:
\begin{equation}
\tilde m^2_{u,d,e}\sim \tilde m^2\left(
 \begin{matrix}
  1 & 0 & 0 \\
  0 & 1+\epsilon^2 & \epsilon \\
  0 & \epsilon & O(1) 
 \end{matrix}
\right).
\end{equation}
Here, $\epsilon^2$ results
from higher
dimensional interactions, like 
$\int d^4\theta(\Psi_a\bar F^a)^\dagger \Psi_b\bar F^bZ^\dagger Z$,
through a non-vanishing VEV $\VEV{\bar F}$. 
The important parameters which are constrained 
by the FCNC processes are
defined by
\begin{equation}
\delta_x\equiv V_x^\dagger\frac{\tilde m^2_x-\tilde m^2}{\tilde m^2}
V_x,
\end{equation}
where $V_x$ is a diagonalizing matrix for fermions 
$x=q,u_R,d_R,l,e_R$\cite{GGMS}.
The constraints are given as, for example,
\begin{eqnarray}
\sqrt{|{\rm Im}(\delta_{D_L})_{12}(\delta_{D_R})_{12})|}&\leq&
 2\times 10^{-4}\left(\frac{\tilde m_Q}{500\ {\rm GeV}}\right) \nonumber \\
|{\rm Im}(\delta_{D_R})_{12}| & \leq & 1.5\times 10^{-3}
\left(\frac{\tilde m_Q}{500\ {\rm GeV}}\right),
\label{K}
\end{eqnarray}
at the weak scale from $\epsilon_K$ in the $K$ meson mixing,
and 
\begin{equation}
|(\delta_{E_L})_{12}|\leq 4\times 10^{-3}\left(
\frac{\tilde m_L}{100\ {\rm GeV}}\right)^2
\label{mu}
\end{equation}
from the $\mu\rightarrow e\gamma$ process.

Actually, the $U(2)_H$ symmetry realizes not only 
hierarchical Yukawa couplings 
but also approximately universal sfermion masses 
of the first two generation fields.
These mass matrices lead to the relations
\begin{equation}
\frac{\tilde m_2^2-\tilde m_1^2}{\tilde m^2}
\sim \frac{m_{F2}}{m_{F3}},
\label{Mass}
\end{equation}
where $m_{Fi}$ and $\tilde m_i$ are the masses of the $i$-th generation fermions
and the $i$-th generation sfermions, respectively.
Unfortunately, these predictions of this simple model imply
a problematic contribution to the
$\epsilon_K$ parameter in the $K$ meson mixing and 
the $\mu\rightarrow e\gamma$ process. 
Moreover, it is obvious that the hierarchical
Yukawa couplings predicted by this simple model are similar for the 
up-quark sector, the down-quark sector, and the charged-lepton sector.
This is inconsistent with the experimental results.
Moreover,  in the neutrino sector, 
it seems to be difficult to obtain 
the large neutrino mixing angles that have been measured in some recent 
experiments \cite{atmos,solar}. 

One of the most natural solutions for the above problems is
to introduce $E_6$ unification. One of the points is that
the Cabibbo-Kobayashi-Maskawa (CKM) mixings are obtained from the 
mixings of the diagonalizing matrix 
of ${\bf 10}$ fields of $SU(5)$ that
includes the doublet quark $Q$ and 
the Maki-Nakagawa-Sakata (MNS) mixings
are obtained from the mixings of 
the diagonalizing matrix of ${\bf \bar 5}$
fields of $SU(5)$ that includes the doublet lepton $L$. 
The fundamental representation ${\bf 27}$ of $E_6$ is divided as
\begin{equation}
{\bf27} \rightarrow \underbrace{
[{\bf 10}_1 +{\bf \bar 5}_{-3}+{\bf 1}_5]}_{{\bf 16}_1}
+\underbrace{[{\bf \bar 5}_2+{\bf 5}_{-2}]}_{{\bf 10}_{-2}}
+ \underbrace{[{\bf 1}_{0}]}_{{{\bf 1}_4}}
\label{27}
\end{equation}
under $E_6\supset SO(10)\times U(1)_{V'}\supset SU(5)\times U(1)_V
\times U(1)_{V'}$. Here,
the representations of $SO(10)\times U(1)_{V'}$ and $SU(5)\times U(1)_V$ 
are explicitly denoted.
Therefore, if we introduce three ${\bf 27}_i$ $(i=1,2,3)$ 
for the three generation
quarks and leptons, 
only three of the six ${\bf \bar 5}$ fields represent 
quarks and leptons in our world and 
the other ${\bf \bar 5}$s become superheavy.
One of the most important assumptions is that the main light modes of 
${\bf \bar 5}$ fields come from the first two generation fields 
${\bf 27}_1$ and ${\bf 27}_2$\cite{Etwist}. 
The assumption that ${\bf\bar 5}$ fields from the third 
generation field ${\bf 27}_3$
has larger couplings, that is, larger masses is natural because the 
third generation field ${\bf 27}_3$ must have larger Yukawa coupling to 
realize the large top quark mass.
Then, actually, the ${\bf \bar 5}$ fields can have 
milder hierarchical Yukawa
couplings than the ${\bf 10}$ fields, and therefore, 
the MNS mixings become
larger than the CKM mixings. 
When we introduce a horizontal symmetry $SU(2)_H$,
all the three generation ${\bf \bar 5}$ fields come 
from a single field
${\bf 27}_a$, which results in universal sfermion masses 
for ${\bf\bar 5}$
fields. Note that the $E_6$ symmetry is sufficient 
but not necessary for the above
structure. The sufficient and necessary symmetry is $SU(2)_E$ 
that rotates
two ${\bf \bar 5}$s and two ${\bf 1}$s in ${\bf 27}$ representation as
 doublets.

\section{$SO(10)\times SU(2)_H$}
In this section, we consider a model with an
$SO(10)\times SU(2)_H\times U(1)_A$
gauge symmetry. 
This model does not solve the SUSY flavor problem sufficiently, 
but it is shown that in the Higgs sector, the horizontal symmetry 
can be naturally introduced.

\subsection{Doublet-triplet splitting in $SO(10)$}
\label{so10review}
In this subsection, we review a mechanism that 
realizes the DT splitting naturally 
in the $SO(10)$ unified scenario
without a non-abelian horizontal gauge symmetry.
\cite{maekawa}

The content of the Higgs sector with the $SO(10)\times U(1)_A$ gauge 
symmetry 
is given in Table \ref{HiggsContentSO10}, 
where the symbols $``+"$ and $``-"$ denote 
$Z_2$ parity quantum numbers.
\begin{table}
\caption{Typical values of anomalous $U(1)_A$ charges. The symbols $\pm$ 
denote a $Z_2$ parity.
}
\label{HiggsContentSO10}
\begin{center}
\begin{tabular}{|c|c|c|} 
\hline
                  &   non-vanishing VEV  & vanishing VEV \\
\hline 
{\bf 45}          &   $A(a=-1,-)$        & $A'(a'=3,-)$      \\
{\bf 16}          &   $C(c=-3,+)$        & $C'(c'=2,-)$      \\
${\bf \overline{16}}$&$\bar C(\bar c=0,+)$ & $\bar C'(\bar c'=5,-)$ \\
{\bf 10}          &   $H(h=-3,+)$        & $H'(h'=4,-)$      \\
{\bf 1}           &$\Theta(\theta=-1,+)$,$Z(z=-2,-)$,$\bar Z(\bar z=-2,-)$& 
                  $S(s=3,+)$ \\
\hline
\end{tabular}
\end{center}
\end{table}

One of the most important features of the GUT scenario with the anomalous
$U(1)_A$ symmetry is that 
VEVs are determined
by anomalous $U(1)_A$ charges as
\begin{equation}
\VEV{O_i}\sim \left\{ 
\begin{array}{ccl}
  \lambda^{-o_i} & \quad & o_i\leq 0 \\
  0              & \quad & o_i>0
\end{array} \right. ,
\label{VEV}
\end{equation}
where the $O_i$ are GUT gauge singlet operators with 
charges
$o_i$, and $\lambda\equiv \VEV{\Theta}/\Lambda\ll 1$.
Here the Froggatt-Nielsen (FN) field $\Theta$ has 
an anomalous $U(1)_A$ charge of $-1$ \cite{FN}. 
Throughout this paper, we denote all superfields and chiral 
operators by uppercase letters and their anomalous $U(1)_A$ 
charges by the corresponding lowercase letters. 
When convenient, we
use units in which $\Lambda=1$. 
 Such a vacuum structure is naturally obtained 
if we introduce generic
interactions even for higher-dimensional operators and 
if the $F$-flatness 
conditions determine the scale of the VEVs\cite{maekawa,BM}. 
Since all the positively charged singlet operators 
have vanishing VEVs,
the SUSY zero (holomorphic zero) mechanism acts effectively. 
Namely, the negatively
charged interactions are not allowed by the symmetry.

Therefore the superpotential for determination of the
VEVs can be written as 
\begin{eqnarray}
W&=&W_{H^\prime}+ W_{A^\prime} + W_S + W_{C^\prime}+W_{\bar C^\prime}, 
\label{W} \\
W_{A^\prime}&=&\lambda^{a^\prime+a} A^\prime A+\lambda^{a^\prime+3a}(
(A^\prime A)_{\bf 1}(A^2)_{\bf 1}
+(A^\prime A)_{\bf 54}(A^2)_{\bf 54}), 
\label{WA} \\
W_S&=&\lambda^{s+c+\bar c}S\left((\bar CC)+\lambda^{-(c+\bar c)}
+\lambda^{-(c+\bar c)+2a}A^{2}\right),
\label{WS} \\
W_{C^\prime}&=&
       \bar C(\lambda^{\bar c^\prime +c+a}A
       +\lambda^{\bar c^\prime +c+\bar z}\bar Z)C^\prime, 
\label{WC} \\
W_{\bar C^\prime}&=&
       \bar C^\prime(\lambda^{\bar c^\prime +c+a} A
       +\lambda^{\bar c^\prime +c+z}Z)C,
\label{WCB}\\
W_{H^\prime}&=&\lambda^{h+a+h^\prime}H^\prime AH. 
\label{WH}
\end{eqnarray}
Here $W_X$ denotes the terms linear in the field $X$ that has 
positive anomalous $U(1)_A$ charge 
and we omit $\order1$ coefficients. 
Note that terms including two
fields with positive charge, like 
$\lambda^{2h^\prime}H^\prime H^\prime$,
give contributions to the mass terms but not to the VEVs.
In Eq. (\ref{WA}), the suffices {\bf 1} and {\bf 54} indicate the 
representations 
of the composite
operators under the $SO(10)$ gauge symmetry.
In the above equations, for simplicity, we ignore
terms like 
${\bf 16}^4$, ${\bf \overline{16}}^4$, ${\bf 10\cdot 16}^2$,
 ${\bf 10 \cdot \overline{16}}^2$ and ${\bf 1\cdot 10}^2$, 
 even if they
are allowed by the symmetry. 
This is because these interactions do not play a significant role
in our argument, because they do not include products of only 
the neutral components under the standard gauge group.
 It is easy to include these terms in
our analysis.

The $F$-flatness condition of $A'$ have a solution in which
$\VEV{A}=i\tau_2\times {\rm diag}(v,v,v,0,0)$, which 
breaks $SO(10)$ into 
$SU(3)_C\times SU(2)_L\times SU(2)_R\times U(1)_{B-L}$.
It is interesting that the scale $v$ is determined by the anomalous 
$U(1)_A$ charge as $v\sim \lambda^{-a}$.
 This Dimopoulos-Wilczek (DW) form of the VEV plays an
important role in solving the DT splitting problem\cite{DW}.

The $F$-flatness condition of $S$ determines
the scale of the VEV $\VEV{\bar C C}$
as $\VEV{\bar CC}\sim \lambda^{-(c+\bar c)}$, and 
the $D$-flatness condition requires 
$|\VEV{C}|=|\VEV{\bar C}|\sim \lambda^{-(c+\bar c)/2}$.
The scale of the VEV is again determined by only the charges of 
$C$ and $\bar C$.

The $F$-flatness conditions of $C^\prime$ and 
$\bar C^\prime$ realize the alignment of the VEVs 
$\VEV{C}$ and $\VEV{\bar C}$ (Barr-Raby mechanism)\cite{BR}.
The $F$-flatness conditions $F_{C^\prime}=F_{\bar C^\prime}=0$ give
$(\lambda^{a-z} A+Z)C=\bar C(\lambda^{a-\bar z} A+\bar Z)=0$. 
Recall that the VEV of $A$ is 
proportional to the $B-L$ generator $Q_{B-L}$ as 
$\VEV{A}=\frac{3}{2}vQ_{B-L}$.
Also $C$, ${\bf 16}$, is decomposed into 
$({\bf 3},{\bf 2},{\bf 1})_{1/3}$, 
$({\bf \bar 3},{\bf 1},{\bf 2})_{-1/3}$, 
$({\bf 1},{\bf 2},{\bf 1})_{-1}$ and $({\bf 1},{\bf 1},{\bf 2})_{1}$ 
under
$SU(3)_C\times SU(2)_L\times SU(2)_R\times U(1)_{B-L}$.
Since $\VEV{\bar CC}\neq 0$, 
not all components in the spinor $C$ vanish. 
Then $Z$ is fixed as $Z\sim -\frac{3}{2}\lambda v Q_{B-L}^0$, 
where $Q_{B-L}^0$ is
the $B-L$ charge of the component field that has 
the non-vanishing VEV in $C$. 
It is interesting
that no other component fields can have non-vanishing VEVs, 
because of the $F$-flatness
conditions. If  the $({\bf 1},{\bf 1},{\bf 2})_1$ field obtains
a non-zero VEV (and therefore $\VEV{Z}\sim -\frac{3}{2}\lambda v$), 
then the gauge group 
$SU(3)_C\times SU(2)_L\times SU(2)_R\times U(1)_{B-L}$ is broken to the 
standard gauge group. Once the direction of the VEV $\VEV{C}$ is 
determined, the VEV $\VEV{\bar C}$ must have the same direction, 
because of the $D$-flatness
condition. Therefore, $\VEV{\bar Z}\sim -\frac{3}{2}\lambda v$.
Thus, all VEVs have now been fixed.

Finally, the $F$-flatness condition of $H^\prime$ is examined. 
The $F$-flatness condition $F_{H^\prime}=0$ leads to the vanishing 
VEV of the triplet Higgs, $\VEV{H_T}=0$. 

Here, we examine the mass spectrum of ${\bf 5}$ and ${\bf \bar 5}$ of
$SU(5)$, to see how the DT splitting is realized.
Considering the additional terms
$H^\prime H^\prime$,
$\bar CA\bar C'H$, $\bar C\bar C'H'$, 
$CC'H'$, $H'\bar C\bar CA$ and $\bar C'C'$
we write the mass matrices $M_I$,
which are for the representations $I=D^c(H_T),L(H_D)$ and their 
conjugates:
\begin{equation}
M_I=\bordermatrix{
       &I_H & I_{H^\prime} & I_{C} &I_{C'} \cr
\bar I_H & 0 & \lambda^{h+h'+a}\VEV{A} & 0 & 0 \cr
\bar I_{H'} & \lambda^{h+h'+a}\VEV{A} & \lambda^{2h'} 
& 0 & \lambda^{h'+c+c'}\VEV{C} \cr
\bar I_{\bar C}& 0 & \lambda^{h'+2\bar c}\VEV{\bar C} & 0 
& \lambda^{\bar c+c'} \cr
\bar I_{\bar C'} &  \lambda^{\bar c+\bar c'+h}\VEV{\bar C} & 
 \lambda^{h'+\bar c+\bar c'}\VEV{\bar C}
 & \lambda^{c+\bar c'} 
  & \lambda^{c'+\bar c'} \cr}.
\end{equation}
In this matrix, $\VEV{A}$ vanishes only for doublet Higgs,
so only one pair of doublet Higgs becomes massless.
Massless modes are
\begin{eqnarray}
H_D&=&{\bf \bar 5}_H +\lambda^{\bar c-c+h}\VEV{\bar C}{\bf \bar 5}_C, \\
\bar H_D&=& {\bf 5}_H.
\end{eqnarray}
Namely the DT splitting is realized.
The effective colored
Higgs mass is estimated as
$(\lambda^{h+h^\prime})^2/\lambda^{2h^\prime}=\lambda^{2h}$, 
which is larger than the cutoff scale, because
$h<0$. Therefore the proton decay via dimension 5 operators
is naturally suppressed.

Straightforward calculation shows that all the non-singlet fields
become massive except one pair of doublet Higgs discussed above.
Surprisingly, though the mass spectrum does not respect 
the $SU(5)$ GUT symmetry, 
the gauge coupling unification can be naturally realized.
This is because the general discussions for the natural gauge coupling 
unification in Refs.\cite{NGCU} can be applied.
This gauge coupling unification requires that
the cutoff scale $\Lambda$ must be around the usual GUT scale 
$\Lambda_G\sim 2\times 10^{16}$ GeV and the unification scale becomes
$\lambda^{-a}\Lambda$, which makes proton decay via dimension 6 operators
interesting.

\subsection{Doublet-triplet splitting in $SO(10)\times SU(2)_H$}
 The content of Higgs sector is listed 
in Table \ref{HiggsContentSO10SU2}.
\begin{table}
\caption{Typical values of anomalous $U(1)_A$ charges.}
\label{HiggsContentSO10SU2}
\begin{center}
\begin{tabular}{|c|c|} 
\hline
$SO(10)\times SU(2)_H$  &    \\
\hline 
({\bf 45,1})          &   $A(a=-1/2)$,\quad $A'(a'=3/2)$      \\
({\bf 1,3})           &   $B(b=-7/2)$            \\
({\bf 16,2})          &   $C(c=-1)$          \\
$({\bf \overline{16},\bar 2})$&$\bar C(\bar c=5/2)$  \\
({\bf 10,1})          &   $H(h=-9/2)$,\quad $H'(h'=5)$      \\
({\bf 1,1})           &$\Theta(\theta=-1)$,\quad $Z(z=-3/2)$,\quad 
                       $S(s=2)$,\quad $S'(s'=13/2)$\\
({\bf 1,2})           &  $F(f=-3/2)$,\quad $F'(f'=2)$ \\
\hline
\end{tabular}
\end{center}
\end{table}
The essential part of the superpotential to determine the vacuum structure
is written as
\begin{eqnarray}
W&=&A'A+A'A^3+\bar C(A+Z)C+H'AH+S(1+\bar CBC) \nonumber \\
 &&+F'(\bar CC)F+S'(1+FBF+\bar CBC+\bar CFCF).
 \label{Potential}
\end{eqnarray}
Here we omit the $\order1$ coefficients as well 
as the power factor $\lambda^x$
where $x$ is the total charge of the corresponding term.
The half integer charges play the same role 
as $Z_2$ parity in the previous
$SO(10)$ model.
The last term in Eq.(\ref{Potential}) 
and fields $F$ and $F'$
are required to obtain realistic quark and lepton mass matrices, 
but not required to realize the DT splitting.
Besides the last term, the structure is essentially the same as in the 
previous models.
The difference appears in the terms $\bar C(A+Z)C$ and $S(1+\bar CBC)$.
Since $c+\bar c>0$, the VEV $\VEV{\bar CC}$ must have vanishing VEV.
On the other hand, the $F$-flatness condition of $S$ leads to 
the non-vanishing VEV $\VEV{\bar CBC}\sim \lambda^{-(\bar c+b+c)}$.
Therefore $\bar C$, $B$ and $C$ must have non-vanishing VEVs. 
If we take
\begin{eqnarray}
\VEV{C}&=&\left(\begin{array}{c}
               0 \\
               V_C 
              \end{array}\right), \quad 
\VEV{\bar C} =\left(V_{\bar C}  \quad 0\right), \\
\VEV{B}&=&\left(\begin{array}{c c}
               0 & V_B \\
               0 & 0 
              \end{array}\right)
\end{eqnarray}
with $|V_C|=|V_{\bar C}|=|V_B|\sim \lambda^{-\frac{1}{3}(c+\bar c+b)}$,
it is easily checked that 
not only the above conditions from $F$-flatness conditions but also the
$D$-flatness conditions for $SO(10)$ and $SU(2)$ gauge interactions
are satisfied.

The $F$-flatness condition of $\bar C$  
\begin{equation}
(A+Z)C=(Q_{B-L}\lambda^a+Z\lambda^z)C=0
\end{equation}
realizes alignment (Barr-Raby mechanism).
As in the previous subsection, The VEV of 
 $Z$ is fixed such that $Z\sim -\frac{3}{2}\lambda v Q_{B-L}^0$, 
where $Q_{B-L}^0$ is
the $B-L$ charge of the component of $C$ that has non-vanishing VEV. 
Note that once we fix the VEV of $Z$, the $F$-flatness condition of 
$\bar C$ automatically realizes the same alignment for the VEV of $C$,
which is consistent with the $D$-flatness condition of $SO(10)$.
Namely, an additional sliding singlet $\bar Z$ does not required.
This interesting feature is realized by unifying the Higgs sectors for
breaking $SU(2)_H$ and $SO(10)$.

Considering the additional terms
$H^\prime H^\prime$,
$\bar C^2AH$, $C^2\bar C^2$, $H'CC$ and
$\bar C^2(1+BA)H'$,
we write the mass matrices of ${\bf 5}$ and ${\bf \bar 5}$ of
$SU(5)$,
which are for the representations $I=D^c(H_T),L(H_D)$ and their 
conjugates:
\begin{equation}
M_I=\bordermatrix{
       &I_H & I_{H^\prime} & I_{C_2} &I_{C_1} \cr
\bar I_H & 0 & \lambda^{h+h'+a}\VEV{A} & 0 & 0 \cr
\bar I_{H'} & \lambda^{h+h'+a}\VEV{A} & \lambda^{2h'} 
& 0 & \lambda^{h'+2c}\VEV{C} \cr
\bar I_{\bar C_1}& 0 & \lambda^{h'+2\bar c+b}\VEV{\bar CB} & 0 
& \lambda^{\bar c+c} \cr
\bar I_{\bar C_2} &  \lambda^{2\bar c+h}\VEV{\bar C} & 
 \lambda^{h'+2\bar c}\VEV{\bar C}
 & \lambda^{c+\bar c} 
  & \lambda^{2c+2\bar c}\VEV{\bar CC} \cr}.
\end{equation}
In this matrix, $\VEV{A}$ vanishes only for doublet Higgs,
so only one pair of doublet Higgs 
\begin{eqnarray}
H_D&=&{\bf \bar 5}_{C_2}
+\frac{\lambda^{c+\bar c}}{\lambda^{2\bar c+h}\VEV{\bar C}}{\bf \bar 5}_H,
 \label{dhiggs}\\
\bar H_D&=& {\bf 5}_H
\label{higgs}
\end{eqnarray}
becomes massless. 
This mixing in the doublet Higgs plays an important role in obtaining
quark and lepton mass matrices.

The $F$-flatness condition of $F'$ leads to 
\begin{equation}
\VEV{F}=\left(\begin{matrix} 0  \\ V_F\end{matrix}\right),\quad 
V_F\sim \lambda^{-f-\frac{1}{3}b+\frac{1}{6}(c+\bar c)},
\end{equation}
where the non-vanishing scale $V_F$ is determined by the $F$-flatness
condition of $S'$ that leads to $\VEV{FBF}\sim \lambda^{-(2f+b)}$.
Again, an alignment happens by the $F$-flatness condition.
This non-vanishing VEV is important in obtaining realistic mass
matrices.

When the VEV relation (\ref{VEV}) holds, 
as discussed in Ref.\cite{NGCU}, the effective charges 
$\tilde x$ of fields $X$
are defined as
\begin{equation}
\tilde x=x+c_F\Delta f+c_V\Delta c
\end{equation}
where $c_F$ and $c_V$ are $U(1)_F \subset SU(2)_F$ 
and $U(1)_V$ charges of $X$ which
are normalized by the charges of the component 
with non-vanishing VEVs 
$\VEV{F}$ and $\VEV{C}$, respectively.
Note that the GUT and flavor singlet operators $O=\prod Y_i$ have the effective 
charges $\tilde o_i=o_i$ because $\sum c_F(Y_i)=\sum c_V(Y_i)=0$.
$\Delta f$ and $\Delta c$ are determined by the relations
\begin{equation}
\VEV{X}\sim \lambda^{-\tilde x},
\end{equation}
where $X={\bf 1}_{\bar C}, {\bf 1}_C, {\bf 1}_F, \cdots$, 
as 
\begin{equation}
\Delta f=\frac{1}{6}(c+\bar c)-\frac{1}{3}b, \quad 
\Delta c=-\frac{1}{2}(c-\bar c).
\end{equation}
It is easily checked that the effective charges defined 
in the above determine
all the scales of non-vanishing VEVs as 
$\VEV{X}\sim \lambda^{-\tilde x}$ 
and masses of superheavy fields of $Y_1$ and $Y_2$ as 
$\lambda^{\tilde y_1+\tilde y_2}$. 
Note that all the terms that reproduce the mass term 
contribute $\order{\lambda^{\tilde y_1+\tilde y_2}}$ to the mass of $Y_1$ 
and $Y_1$.
For example, 
$\lambda^{y_1+y_2+\sum x_i}Y_1Y_2\prod X_i=\lambda^{\tilde y_1+\tilde y_2+\sum
\tilde x_i}Y_1Y_2\prod X_i$ give $\lambda^{\tilde y_1+\tilde y_2}Y_1Y_2$ by
developing the VEVs $\VEV{X_i}\sim \lambda^{-\tilde x_i}$.
It is obvious that even for higer dimensional interactions, the coefficients 
are determined by the simple sums of the effective charges.
Thus, in the followings, we often write down only some representatives of
interactions.

\subsection{Quark and lepton sector}
It is not so difficult to obtain realistic quark and lepton mass matrices,
using this simple Higgs sector.
For example, 
let us introduce the matter sector as in Table \ref{MatterContentSO10SU2}.
\begin{table}
\caption{The typical values of anomalous $U(1)_A$ charges for quarks and 
leptons are listed.
The odd quarter integer charges play the same role as odd $R$-parity.}
\label{MatterContentSO10SU2}
\begin{center}
\begin{tabular}{|c|c|c|c|c|} 
\hline
   &   $\Psi$ & $\Psi_3$ & $T$ \\
\hline 
 $SO(10)$ & {\bf 16} & {\bf 16} & {\bf 10}  \\
 $SU(2)_H$ & {\bf 2} &{\bf 1} & {\bf 1}   \\
 $U(1)_A$ & $\psi=29/4$ & $\psi_3=9/4$ & $t=15/4$  \\
 \hline
\end{tabular}
\end{center}
\end{table}
Here the rational charges play the same role as $R$-parity.

The {\bf 10} field is divided into ${\bf 5}+{\bf \bar 5}$ of $SU(5)$,
thus one pair of ${\bf 5}+{\bf \bar 5}$ becomes massive.
First of all, we examine which modes become massless.
The mass matrix $M_I$ of ${\bf 5}$ and ${\bf \bar 5}$ are determined
by the effective charges
\begin{eqnarray}
\tilde \psi_1({\bf \bar 5})&\equiv&\psi-\frac{3}{5}\Delta c+\Delta f
        =\frac{29}{4}+\frac{11}{30} \\
\tilde \psi_2({\bf \bar 5})&\equiv&\psi-\frac{3}{5}\Delta c-\Delta f
        =\frac{29}{4}-\frac{37}{15} \\
\tilde \psi_3({\bf \bar 5})&\equiv&\psi_3-\frac{3}{5}\Delta c
        =\frac{9}{4}-\frac{21}{20} \\
\tilde t({\bf \bar 5})&\equiv&t+\frac{2}{5}\Delta c
        =\frac{15}{4}+\frac{7}{10} \\
\tilde t({\bf 5})&\equiv&t-\frac{2}{5}\Delta c
        =\frac{15}{4}-\frac{7}{10} 
\end{eqnarray}
from the interactions
\begin{equation}
W_M=\Psi_3TBCF+T\Psi(1+B(A+Z))C+T^2
\end{equation}
as
\begin{equation}
M_I=
\bordermatrix{
       & I_{\Psi 3} & I_{\Psi 2} & I_{\Psi 1}  & I_T \cr
\bar I_T & \lambda^{\tilde t({\bf 5})+\tilde \psi_3({\bf \bar 5})} & 
           \lambda^{\tilde t({\bf 5})+\tilde \psi_2({\bf \bar 5})} & 
           \lambda^{\tilde t({\bf 5})+\tilde \psi_1({\bf \bar 5})} &
           \lambda^{\tilde t({\bf 5})+\tilde t({\bf \bar 5})} \cr},
\end{equation}
where $I=D^c, L$. 
Here, we do not write down all the interactions that produce the elements
of the above matrix, because their contributions have the same power of 
$\lambda$.
Thus, the light modes become
\begin{eqnarray}
I_1&\sim&{\bf \bar 5}_{\psi 1}
+\lambda^{\tilde \psi_1({\bf \bar 5})-\tilde \psi_3({\bf \bar 5})}
{\bf \bar 5}_{\Psi 3},\nonumber \\
I_2&\sim&{\bf \bar 5}_{\psi 2}+\lambda^{\tilde \psi_2({\bf \bar 5})
-\tilde \psi_3({\bf \bar 5})}{\bf \bar 5}_{\Psi 3}, \label{matter}\\
I_3&\sim&{\bf \bar 5}_{T}+\lambda^{\tilde t({\bf \bar 5})
-\tilde \psi_3({\bf \bar 5})}{\bf \bar 5}_{\psi 3}
\nonumber
\end{eqnarray}
in a certain base of light ${\bf \bar 5}$ fields\footnote{
Note that if $F$ has a vanishing VEV, the $\bar 5_{\Psi 3}$ becomes one of
the light modes, that leads to unrealistic quark mass matrices.}.
Yukawa couplings of quarks and leptons are obtained from the 
superpotential
\begin{eqnarray}
W_Y&=&\Psi_3^2H+\Psi_3\Psi F(A+B)H+\Psi^2(AZ+F^2+AB)H 
\nonumber \\
&& +T\Psi_3CF+T\Psi(1+AB)C
\end{eqnarray}
as
\begin{eqnarray}
Y_u&\sim & \lambda^{\tilde h({\bf 5})}\left(
 \begin{matrix}
  \lambda^{\tilde\psi_1({\bf 10})+\tilde\psi_1({\bf 10})} & 
  \lambda^{\tilde\psi_1({\bf 10})+\tilde\psi_2({\bf 10})} 
  & \lambda^{\tilde\psi_1({\bf 10})+\tilde\psi_3({\bf 10})} 
 \\
  \lambda^{\tilde\psi_1({\bf 10})+\tilde\psi_2({\bf 10})} 
 & \lambda^{\tilde\psi_2({\bf 10})+\tilde\psi_2({\bf 10})}
   & \lambda^{\tilde\psi_2({\bf 10})+\tilde\psi_3({\bf 10})} \\
   \lambda^{\tilde\psi_1({\bf 10})+\tilde\psi_3({\bf 10})} & 
   \lambda^{\tilde\psi_2({\bf 10})+\tilde\psi_3({\bf 10})} & 
   \lambda^{\tilde\psi_3({\bf 10})+\tilde\psi_3({\bf 10})}
 \end{matrix}\right), \\
Y_d&\sim & Y_e^T\sim\lambda^{\tilde c_2({\bf \bar 5})}\left(
 \begin{matrix}
     \lambda^{\tilde\psi_1({\bf 10})+\tilde\psi_1({\bf \bar 5})} & 
     \lambda^{\tilde\psi_1({\bf 10})+\tilde\psi_2({\bf \bar 5})} 
 & \lambda^{\tilde\psi_1({\bf 10})+\tilde t({\bf \bar 5})} \\
     \lambda^{\tilde\psi_2({\bf 10})+\tilde\psi_1({\bf \bar 5})} 
 &   \lambda^{\tilde\psi_2({\bf 10})+\tilde\psi_2({\bf \bar 5})} 
 &  \lambda^{\tilde\psi_2({\bf 10})+\tilde t({\bf \bar 5})}  \\
 \lambda^{\tilde\psi_3({\bf 10})+\tilde\psi_1({\bf \bar 5})} & 
 \lambda^{\tilde \psi_3({\bf 10})+\tilde\psi_2({\bf \bar 5})} 
   & \lambda^{\tilde\psi_3({\bf 10})+\tilde t({\bf \bar 5})} 
 \end{matrix}
                   \right), \\
Y_{\nu_D}^T&\sim &\lambda^{\tilde h({\bf 5})}\left(
 \begin{matrix}
     \lambda^{\tilde\psi_1({\bf 1})+\tilde\psi_1({\bf \bar 5})} 
      & 
     \lambda^{\tilde\psi_1({\bf 1})+\tilde\psi_2({\bf \bar 5})} 
 &      \lambda^{\tilde\psi_1({\bf 1})+\tilde t({\bf \bar 5})} \\
     \lambda^{\tilde\psi_2({\bf 1})+\tilde\psi_1({\bf \bar 5})} 
 &   \lambda^{\tilde\psi_2({\bf 1})+\tilde\psi_2({\bf \bar 5})} 
 &   \lambda^{\tilde\psi_2({\bf 1})+\tilde t({\bf \bar 5})}  \\
 \lambda^{\tilde\psi_3({\bf 1})+\tilde\psi_1({\bf \bar 5})} & 
 \lambda^{\tilde \psi_3({\bf 1})+\tilde\psi_2({\bf \bar 5})} 
   & \lambda^{\tilde\psi_3({\bf 1})+\tilde t({\bf \bar 5})} 
 \end{matrix}
                   \right). 
\end{eqnarray}
                 
Note that down-quark and charged lepton Yukawa matrices have additional
contribution given through the Higgs mixing in Eq. (\ref{dhiggs}), which
is the same order as that from the Higgs ${\bf\bar 5}_H$.
This is guaranteed by the fact that the matter mixings in Eq. (\ref{matter})
and the Higgs mixing are determined by the difference of the effective charges.
For example, $\lambda^{\psi+t+c}\Psi T C$ and 
$\lambda^{\psi+\psi_3+h+f}\Psi F\Psi_3H$ give Yukawa interactions 
$\lambda^{\tilde \psi_1({\bf 10})+\tilde t({\bf \bar 5})
+\tilde c_2({\bf \bar 5})}
{\bf 10}_{\Psi_1}{\bf \bar 5}_T{\bf \bar 5}_{C_2}$
and 
$\lambda^{\tilde \psi_1({\bf 10})+\tilde \psi_3({\bf \bar 5})
+\tilde h({\bf \bar 5})}
{\bf 10}_{\Psi_1}{\bf \bar 5}_{\Psi_3}{\bf \bar 5}_H$,
respectively, and the ratio of the Yukawa couplings, 
$\lambda^{\tilde t({\bf \bar 5})+\tilde c_2({\bf \bar 5})
-\tilde \psi_3({\bf \bar 5})-\tilde h({\bf \bar 5})}$ is nothing but the
product of the mixing coefficients in Eqs. (\ref{dhiggs}) and (\ref{matter}).
As a result, CKM mixings can be obtained as
$V_{ij}\sim \lambda^{|\tilde \psi_i({\bf 10})-\tilde \psi_j({\bf 10})|}$.

Because The mass matrix of neutrinos is straightforwardly calculated.
From the interactions
\begin{equation}
W=(\Psi_3\Psi_3+\Psi_3\Psi(A+B) F +(\Psi(A+B) F)^2)\bar C^2B
\end{equation}
we obtain the right-handed neutrino mass matrix as
\begin{equation}
(M_{\nu_R})_{ij}\sim \lambda^{\tilde \psi_i({\bf 1})+\tilde \psi_j({\bf 1})}.
\end{equation}
Therefore, light neutrino mass matrix can be estimated from the seesaw relation
as
\begin{equation}
M_{\nu}=Y_{\nu_D}M_{\nu_R}^{-1}Y_{\nu_D}^T \frac{\VEV{h_u}^2\eta^2}{\Lambda}
\sim \lambda^{2\tilde h({\bf 5})}\left(
 \begin{matrix}
   \lambda^{2\tilde \psi_1({\bf \bar 5})} &
   \lambda^{\tilde \psi_1({\bf \bar 5})+\tilde \psi_2({\bf \bar 5})} &
   \lambda^{\tilde \psi_1({\bf \bar 5})+\tilde t({\bf \bar 5})} \\
   \lambda^{\tilde \psi_2({\bf \bar 5})+\tilde \psi_1({\bf \bar 5})} &
   \lambda^{2\tilde \psi_2({\bf \bar 5})} &
   \lambda^{\tilde \psi_2({\bf \bar 5})+\tilde t({\bf \bar 5})} \\
   \lambda^{\tilde t({\bf \bar 5})+\tilde \psi_1({\bf \bar 5})} &
   \lambda^{\tilde t({\bf \bar 5})+\tilde \psi_2({\bf \bar 5})} &
   \lambda^{2\tilde t({\bf \bar 5})} 
 \end{matrix}
   \right)\frac{\VEV{h_u}^2\eta^2}{\Lambda},
\end{equation}
where $\eta$ is a renormalization factor.
Therefore, Maki-Nakagawa-Sakata matrix is obtained as
\begin{equation}
V_{MNS}\sim \left(
 \begin{matrix} 1 &
 \lambda^{\tilde\psi_1({\bf\bar 5})-\tilde\psi_2({\bf \bar 5})} &
 \lambda^{\tilde\psi_1({\bf\bar 5})-\tilde t({\bf \bar 5})} \\ 
 \lambda^{\tilde\psi_1({\bf\bar 5})-\tilde\psi_2({\bf \bar 5})} & 1 &
 \lambda^{\tilde\psi_2({\bf\bar 5})-\tilde t({\bf \bar 5})} \\ 
 \lambda^{\tilde\psi_1({\bf\bar 5})-\tilde t({\bf \bar 5})} &
 \lambda^{\tilde\psi_2({\bf\bar 5})-\tilde t({\bf \bar 5})} & 1 
 \end{matrix}
 \right).
\end{equation}

The concrete model in Tables 2 and 3 gives
\begin{eqnarray}
Y_u&\sim & \left(
 \begin{matrix}\lambda^{\frac{77}{6}} & 
  \lambda^{10} 
  & \lambda^{\frac{77}{12}} \\
  \lambda^{10} 
 & \lambda^{\frac{43}{6}}
   & \lambda^{\frac{43}{12}} \\ 
                \lambda^{\frac{77}{12}} & 
   \lambda^{\frac{43}{12}} & 
   1
 \end{matrix}\right),\ 
Y_d\sim  Y_e^T\sim\left(\begin{matrix}
   \lambda^{\frac{79}{6}} & 
     \lambda^{\frac{31}{3}} 
 & \lambda^{10} \\ 
     \lambda^{\frac{31}{3}} 
 &   \lambda^{\frac{15}{2}} 
 &  \lambda^{\frac{43}{6}}  \\ 
 \lambda^{\frac{27}{4}} & 
 \lambda^{\frac{47}{12}}
   & \lambda^{\frac{43}{12}} 
     \end{matrix}\right), \\
M_{\nu}
&\sim& \lambda^{-\frac{3}{2}}\left(\begin{matrix}
   \lambda^{\frac{19}{3}} &
   \lambda^{\frac{7}{2}} &
   \lambda^{\frac{19}{6}} \\ 
   \lambda^{\frac{7}{2}} &
   \lambda^{\frac{2}{3}} &
   \lambda^{\frac{1}{3}} \\ 
   \lambda^{\frac{19}{6}} &
   \lambda^{\frac{1}{3}} &
   1 \end{matrix}
   \right)\frac{\VEV{h_u}^2\eta^2}{\Lambda},
\end{eqnarray}
from which the CKM matrix and MNS matrix are obtained by
\begin{equation}
V_{CKM}\sim \left(\begin{matrix} 1 &
 \lambda^{\frac{17}{6}} &
 \lambda^{\frac{77}{12}} \\ 
 \lambda^{\frac{17}{6}} & 1 &
 \lambda^{\frac{43}{12}} \\ 
 \lambda^{\frac{77}{12}} &
 \lambda^{\frac{43}{12}} & 1 \end{matrix}
 \right),
V_{MNS}\sim \left(\begin{matrix} 1 &
 \lambda^{\frac{17}{6}} &
 \lambda^{\frac{19}{6}} \\ 
 \lambda^{\frac{17}{6}} & 1 &
 \lambda^{\frac{1}{3}} \\ 
 \lambda^{\frac{19}{6}} &
 \lambda^{\frac{1}{3}} & 1 
\end{matrix}
 \right).
\end{equation}
Taking $\lambda\sim 0.5$ gives reasonable values 
for quark and lepton masses and mixings.
This leads to a bit small solar neutrino mixing angle, 
but a cancellation may make it larger, 
and the bi-large neutrino mixings can be obtained.
 
Unfortunately, the FCNC processes are not 
sufficiently suppressed in this model. 
Actually, the parameters $\delta_{\bf 10}$ and $\delta_{\bf\bar 5}$ 
are estimated as
\begin{eqnarray}
\delta_{\bf 10}&\sim& V_{CKM}^\dagger \left(\begin{matrix}
  \lambda^\frac{4}{3} & \lambda^{\frac{25}{6}} & \lambda^{\frac{77}{12}}\\ 
  \lambda^{\frac{25}{6}} & \lambda^{\frac{4}{3}} & \lambda^{\frac{113}{12}}\\ 
  \lambda^{\frac{77}{12}}&\lambda^{\frac{113}{12}}& 1 
\end{matrix}\right)V_{CKM}
  \sim\left(\begin{matrix}\lambda^\frac{4}{3} &\lambda^\frac{25}{6}
  &\lambda^\frac{77}{12}\\ 
  \lambda^\frac{25}{6}&\lambda^\frac{4}{3}&\lambda^\frac{43}{12}\\ 
  \lambda^\frac{77}{12}&\lambda^\frac{43}{12}&1
\end{matrix}\right), \\
\delta_{\bf\bar 5}&\sim& V_{MNS}^\dagger \left(\begin{matrix}
  \lambda^\frac{4}{3} & \lambda^{\frac{17}{6}} & \lambda^{\frac{11}{3}}\\ 
  \lambda^{\frac{17}{6}} & \lambda^{\frac{4}{3}} & \lambda^{\frac{13}{6}}\\ 
  \lambda^{\frac{11}{3}}&\lambda^{\frac{13}{6}}& 1  
\end{matrix}\right)V_{MNS}\sim 
\left(\begin{matrix}
  \lambda^\frac{4}{3} &\lambda^\frac{17}{6}&\lambda^\frac{19}{6}\\ 
  \lambda^\frac{17}{6}&\lambda^\frac{2}{3}&\lambda^\frac{1}{3}\\ 
  \lambda^\frac{19}{6}&\lambda^\frac{1}{3}&1
 \end{matrix}\right),
\end{eqnarray}
which lead to too large FCNC processes when $\lambda\sim 0.5$.
This is mainly because large mixings in $V_{MNS}$ transform the differences
$\tilde m_3-\tilde m_i$ ($i=1,2$) to the other mixings in the paramters
$\delta_{\bf bar 5}$ and because the $SU(2)_H$ breaking scale,
$|V_C|=|V_{\bar C}|=|V_B|\sim \lambda^{\frac{2}{3}}\sim (0.22)^{\frac{1}{3}}$,
 is not so small to suppress the FCNC processes.


\section{horizontal symmetry in $E_6$ Higgs sectors}

In this section, we investigate $E_6$ models with an anomalous
$U(1)_A$ gauge symmetry whose Higgs 
 sectors have non-trivial quantum numbers of the horizontal 
 symmetry, $SU(2)_H$ or $SU(3)_H$.
Since $E_6$ contains $SU(2)_E$, these model may realize 
 well-suppressed FCNC processes as suggested in \S\ref{horizreview}.
In this sence, $E_6$ models are more promising than 
 $SO(10)$ model considered in the previous section.
Unfortunately, however, if both $E_6$ and the horizontal symmetry
is simultaneously broken, it is anticipated to be difficult to obtain 
realistic models in which the FCNC processes are sufficiently 
suppressed. 
The point is as follows.
In order to suppress sufficiently the FCNC processes by
the horizontal symmetry, 
the scale at which the horizontal symmetry is broken should be smaller
than $\lambda^2$. ( In this section, we take 
$\lambda\sim \sin\theta_C\sim 0.22$, and we do not fix the anomalous $U(1)_A$
charge of the FN field to -1 but $\VEV{\Theta}\sim \lambda^\theta$.)
Generically, 
In GUT models with an anomalous $U(1)_A$ symmetry, it is difficult to
obtain a smaller $E_6$ breaking scale than $\lambda^2$ \cite{E6,reduced}.
Therefore, if both $E_6$ and the horizontal symmetry are broken by a VEV
of a single field, that is, both the symmetries are broken at the same scale,
then the suppression of the FCNC processes does not become sufficient.

After a brief review of the $E_6$ models 
 with an anomalous $U(1)_A$ symmetry and without a horizontal symmetry 
 proposed before, 
 we examine the possibility of the $SU(2)_H$ horizontal symmetry 
 in \S\ref{e6su2} and of the $SU(3)_H$ horizontal symmetry 
 in \S\ref{e6su3}.

\subsection{$E_6$ model without horizontal symmetry}
\label{e6review}

We have proposed two types of $E_6$ Higgs sector 
 in Refs.\cite{E6} and \cite{reduced}. 
A typical charge assignment for each model is displayed 
 in Table \ref{HiggsContentE6} and Table \ref{HiggsContentreduced}, 
respectively. 
\begin{table}
\caption{
The Higgs content of $E_6\times U(1)_A$ model of Ref.\cite{E6}: 
The symbols $\pm$ denote  an additional $Z_2$ parity.
Here, the composite operator $\bar\Phi\Phi$ plays the role of 
 the FN field $\Theta$.}
\label{HiggsContentE6}
\begin{center}
\begin{tabular}{|c|c|c|} 
\hline
                &    \mbox{non-vanishing VEV}  &  \mbox{vanishing VEV} \\
\hline 
{\bf 78}        &   $A(a=-1,-)$        & $A'(a'=4,-)$      \\
{\bf 27}        &   $\Phi(\phi=-3,+)$\  $C(c=-6,+)$ &  $C'(c'=7,-)$  \\
${\bf \cc{27}}$ & $\bar \Phi(\bar \phi=2,+)$ \  $\bar C(\bar c=-2,+)$ 
                & $\bar C'(\bar c'=8,-)$ \\
{\bf 1}         &   $Z_i(z_i=-2,-)$\ $(i=1,2,3)$ \   &  \\
\hline
\end{tabular}
\end{center}
\end{table}
\begin{table}
\caption{
The Higgs content of $E_6\times U(1)_A$ model of Ref.\cite{reduced}: 
The symbols $\pm$ denote an
 additional $Z_2$ symmetry.}
\label{HiggsContentreduced}
\begin{center}
\begin{tabular}{|c|c|c|} 
\hline
                &    \mbox{non-vanishing VEV}  &  \mbox{vanishing VEV} \\
\hline 
{\bf 78}        &   $A(a=-1,+)$        & $A'(a'=5,+)$      \\
{\bf 27}        &   $\Phi(\phi=-3,+)$  &  $C'(c'=6,-)$  \\
${\bf \cc{27}}$ & $\bar C(\bar c=0,-)$ 
                & $\bar\Phi'(\bar c'=5,+)$ \\
{\bf 1}         &   $\Theta(\theta=-1,+)$\ $Z_i(z_i=-1,-)$\ $(i=1,2)$ & \\
\hline
\end{tabular}
\end{center}
\end{table}
The non-vanishing VEVs are given as 
\beqn
 \VEV{\bf{45}_A}&\sim&\lambda^{-a}, \\
 \VEV{\bf{1}_\Phi}=\VEV{\bf{1}_{\bar\Phi}} 
   &\sim& \lambda^{-\half(\phi+\bar\phi)}, \\
 \VEV{\bf{16}_C}=\VEV{\bf{\cc{16}}_{\bar C}}
   &\sim& \lambda^{-\half(c+\bar c)}, 
\eeqn
in the former model and 
\beqn
  \VEV{{\bf45}_A} &\sim& \lambda^{-a} ,\\
  \VEV{{\bf\cc{16}}_{\bar C}} 
    \sim \VEV{{\bf16}_A}
    \sim \VEV{{\bf1}_\Phi}
    &\sim& \lambda^{-\frac{1}{3}(\bar c+a+\phi)}, \\
  \VEV{{\bf{1}}_{\bar C}}
    \sim \VEV{{\bf16}_\Phi}
    &\sim& \lambda^{-\frac{1}{3}(-a+2\bar c+2\phi)} , 
\eeqn
in the latter model.
The correspondence to the $SO(10)$ model in \S\ref{so10review} 
 is understood by the decomposition of $E_6$ representations, 
\beqn
 \bf{78}&\rightarrow&\bf{45}+\bf1+\bf{16}+\bf{\cc{16}} , \\
 \bf{27}&\rightarrow&\bf{16}+\bf{10}+\bf1, 
\eeqn
in terms of $SO(10)$ representations.
The VEVs along the $\bf1$ components of $\bf{27}$ Higgs 
 break $E_6$ into $SO(10)$.
The adjoint Higgs, $A$, of the $SO(10)$ model is embedded into 
 the adjoint Higgs, $A$, in each model.
The spinor Higgs, $C(\bar C)$, is embedded into the field that 
 have an non-vanishing VEV along the $\bf{16}(\bf{\cc{16}})$ 
 component. 
And the MSSM doublet Higgs are embedded into $\Phi$ mainly.

As mentioned below, we use the former type 
 to unify the Higgs sectors for $E_6$ breaking and 
 for horizontal symmetry breaking.
Thus let us concentrate on the former type. 
Here, we also assume the VEV relations Eq.(\ref{VEV}) 
 and thus effective charge is well-defined as
\begin{equation}
\tilde x=x+c_V\Delta c+c_{V'}\Delta\phi,
\end{equation}
where $c_V$ and $c_{V'}$ are $U(1)_V$ and $U(1)_{V'}$ charges of
the field $X$ which are normalized by the charges of the component
with non-vanishing VEVs $\VEV{C}$ and $\VEV{\Phi}$, respectively.
$\Delta c$ and $\Delta\phi$ are determined by the relations
\begin{equation}
\VEV{C}\sim \lambda^{-\tilde c},\quad \VEV{\Phi}\sim \lambda^{-\tilde \phi}.
\end{equation}
This assumption on the VEV relations (\ref{VEV}) is natural if we introduce 
 the generic interaction 
 as comented in \S\ref{so10review}. 
Then, the non-vanishing VEVs are determined 
 by the superpotential 
 that is linear in the field with vanishing VEV, 
\begin{equation}
  W=W_{A^\prime} + W_{C^\prime}+W_{\bar C^\prime}.
\label{e6W}
\end{equation}
$W_{C^\prime}$ and $W_{\bar C^\prime}$ play similar roles 
as in the $SO(10)$ model.
In $W_{A'}$, we have to introduce a term $\bar\Phi A'A^3\Phi$ 
 or $A'A^5$, since the term $A'A^3$ does not contain the term, 
 $({\bf{45}}_{A'}{\bf{45}}_A)_{\bf{54}}({\bf{45}}_{A}^2)_{\bf{54}}$,
 which is required to obtain the DW form of ${\bf{45}}_{A}$ 
 in a natural way. 
And the term, $\bar\Phi A'A\Phi$, is also required, 
 because without this term, some components of $A$ can not obtain 
 superheavy masses due to an accidental symmetry 
 in the superpotential that determine the VEVs, 
 $\VEV A$, $\VEV\Phi$ and $\VEV{\bar\Phi}$, 
 namely they become Pseudo Nambu-Goldstone (PNG) modes. 
However, the larger $a'+a$ results in the lighter mass spectrum 
 for superheavy modes which make the gauge couplings evaluated 
 at the cutoff scale stronger. 
If we take $a'+a>4$, the gauge couplings 
 go to the non-perturbative region 
 in most part of the parameter space. 
Thus, we cannot take $\phi+\bar\phi$ so small, and therefore 
$\VEV\Phi$ and $\VEV{\bar\Phi}$ cannot be so small.
As for the mass spectrum of this Higgs sector, 
 we need the term, $\bar\Phi\bar\Phi\bar C$, to give 
 the effective mass term, 
 $\VEV{{\bf1}_{\bar\Phi}}{\bf{10}}_{\bar\Phi}{\bf{10}}_{\bar C}$. 
This term is essential to avoid unwanted light modes 
 because one pair of $\bf{27}$-$\bf{\cc{27}}$ Higgs 
 can not obtain mass with the primed field in this Higgs content 
 while one pair of $\bf{16}$-$\bf{\cc{16}}$ components 
 can obtain mass of the symmetry breaking scale 
 through the Higgs mechanism.

In the quark and lepton sector, 
 we introduce three fundamental matters, $\Psi_i({\bf 27})\,(i=1,2,3)$, 
 and their $U(1)_A$ charges are taken as $\psi_1-\psi_2\sim 1$ and
 $\psi_2-\psi_3\sim 2$ to reproduce the CKM mixings. 
The three fields, $\Psi_i$, include three $\bf5$s and six $\bf{\bar5}$s of 
$SU(5)$. 
They have mass terms through the two types of interaction 
 $\lambda^{\psi_i+\psi_j+\phi}\Psi_i\Psi_j\Phi$ 
 and $\lambda^{\psi_i+\psi_j+c}\Psi_i\Psi_jC$, 
 yielding a $3\times 6$ mass matrix,
\begin{equation}
  \bordermatrix{
      & {\bf 10}_{\Psi_1} & {\bf 10}_{\Psi_2}
        & {\bf 10}_{\Psi_3} &\quad
      & {\bf 16}_{\Psi_1} & {\bf 16}_{\Psi_2}
        & {\bf 16}_{\Psi_3}\cr
    {\bf 10}_{\Psi_1} 
      & \lambda^{2\psi_1} & \lambda^{\psi_1+\psi_2}
        & \lambda^{\psi_1+\psi_3} &\quad
      & \lambda^{2\psi_1+r} &  \lambda^{\psi_1+\psi_2+r}
        &\lambda^{\psi_1+\psi_3+r}  \cr
    {\bf 10}_{\Psi_2}
      & \lambda^{\psi_1+\psi_2} &\lambda^{2\psi_2}
        & \lambda^{\psi_2+\psi_3} &\quad
      & \lambda^{\psi_1+\psi_2+r}   &  \lambda^{2\psi_2+r}
        &\lambda^{\psi_2+\psi_3+r} \cr
    {\bf 10}_{\Psi_3} 
      & \lambda^{\psi_1+\psi_3} & \lambda^{\psi_2+\psi_3}
        &\lambda^{2\psi_3} &\quad
      & \lambda^{\psi_1+\psi_3+r} & \lambda^{\psi_2+\psi_3+r}
        & \lambda^{2\psi_3+r} \cr  
    }\lambda^{\phi}\VEV{{\bf 1}_{\Phi}}. 
\label{E6-mixing}
\end{equation}
Here, for simplicity, we assume that 
 the SUSY-zero mechanism does not forbit any elements.
However, even if  some of the elements of $3\times 3$ matrix of 
${\bf 10}_{\Psi_i}$ and ${\bf 16}_{\Psi_i}$ vanish by 
the SUSY-zero mechanism,
the following arguments can be applied unless the ranks of 
the $3\times 3$ matrix is reduced.
Here, the parameter $r$ is defined by
 $\lambda^r\equiv\frac{\lambda^\phi\VEV\Phi}{\lambda^c\VEV C}$. 
These mass terms make three pairs of $\bf5$ and $\bf{\bar5}$ 
 superheavy, while three $\bf{\bar5}$s remain massless.
Providing $0\leq r\leq\psi_1-\psi_3$, 
 these three modes can be written as
\begin{alignat}{1}
  {\bf \cc 5}_1 &\sim {\bf 16}_{\Psi_1}
    +\lambda^{\psi_1-\psi_3}{\bf 16}_{\Psi_3}
    +\lambda^{\psi_1-\psi_2+r}{\bf 10}_{\Psi_2}
    +\lambda^{\psi_1-\psi_3+r}{\bf 10}_{\Psi_3}, 
\label{51} \\
  {\bf \cc 5}_2 &\sim {\bf 10}_{\Psi_1}
    +\lambda^{\psi_1-\psi_3-r}{\bf 16}_{\Psi_3}
    +\lambda^{\psi_1-\psi_2}{\bf 10}_{\Psi_2}
    +\lambda^{\psi_1-\psi_3}{\bf 10}_{\Psi_3}, 
\label{52} \\
  {\bf \cc 5}_3 &\sim {\bf 16}_{\Psi_2}
    +\lambda^{\psi_2-\psi_3}{\bf 16}_{\Psi_3}
    +\lambda^{r}{\bf 10}_{\Psi_2}
    +\lambda^{\psi_2-\psi_3+r}{\bf 10}_{\Psi_3},
\label{53}
\end{alignat}
where we use a base in which each light mode includes no 
other main modes.
From these mixing, we can find the Yukawa matrices of 
 the down-type quarks and the charged leptons are given as 
\bequ
Y_d\sim  Y_e^T\sim\left(
\begin{array}{ccc}
      \lambda^{2\psi_1-\psi_2-\psi_3} 
   &  \Ll\lambda^{2\psi_1-\psi_2-\psi_3-r}\Rl
   &  \Ll\lambda^{\psi_1-\psi_3}\Rl \\
      \Ll\lambda^{\psi_1-\psi_3}\Rl &   \Ll\lambda^{\psi_1-\psi_3-r}\Rl 
   &  \lambda^{\psi_2-\psi_3} \\
      \lambda^{\psi_1-\psi_2} & \lambda^{\psi_1-\psi_2-r} & 1 
\end{array}
\Rs\lambda^{\psi_2-\psi_3}, 
\label{5barYukawa}
\eequ
if we set $2\psi_3+\phi=0$ to reproduce 
 the $\order1$ top Yukawa coupling. 
Note that although in the $3\times 6$ matrix (\ref{E6-mixing}) the $SU(2)_R$ 
breaking VEV, $\VEV{C}$, appears through the Yukawa interaction, 
$\Psi_i\Psi_jC$,
this breaking effect is not sufficient to produce the Cabibbo mixing. 
Actually, unless there is  the $SU(2)_{R}$ breaking in the Yukawa couplings
of $\Psi_i\Psi_j\Phi$ and $\Psi_i\Psi_jC$ or in MSSM higgs mixings as in 
$SO(10)$ case (\ref{dhiggs}),
 the components with the parenthesis in the matrix (\ref{5barYukawa}) 
 become smaller 
 in the basis where the Yukawa matrix of the up-type quarks 
 is diagonalized. 
Then, the Cabibbo angle becomes smaller than the naively expected 
 value, $\lambda^{\psi_1-\psi_2}$. 
Therefore, the $SU(2)_{R}$ breaking effect have to appear at least either 
of in the Yukawa couplings or in the MSSM higgs mixings. 
Then, the CKM matrix is obtained as 
\bequ
V_{\rm{CKM}}\sim \Ls
\begin{array}{ccc}
      1   &  \lambda^{\psi_1-\psi_2}  &  \lambda^{\psi_1-\psi_3}\\
      \lambda^{\psi_1-\psi_2} & 1   &  \lambda^{\psi_2-\psi_3} \\
      \lambda^{\psi_1-\psi_3} & \lambda^{\psi_2-\psi_3} & 1 
\end{array}
\Rs\sim
\Ls
\begin{array}{ccc}
      1   &  \lambda  &  \lambda^3\\
      \lambda & 1   &  \lambda^2 \\
      \lambda^3 & \lambda^2 & 1 
\end{array}\Rs.
\label{e6CKM}
\eequ
Thus, the CKM matrix are determined by the differnce of the $U(1)_A$ charges,
$\psi_i-\psi_j$, and therefore, by the difference of the effective $U(1)_A$
charges, $\tilde \psi_i({\bf 16,10})-\tilde \psi_j({\bf 16,10})=\psi_i-\psi_j$.
Here, we denote the representations of $SO(10)$ and $SU(5)$, explicitly.
The MNS matrix is also determined by the difference of the effective charges,
$\tilde \psi_1({\bf 16,\bar 5})-\tilde \psi_1({\bf 10,\bar 5})=r$ and
$\tilde \psi_1({\bf 16,\bar 5})-\tilde \psi_2({\bf 16,\bar 5})=\psi_1-\psi_2$ 
as
\bequ
V_{\rm{MNS}}\sim \Ls
\begin{array}{ccc}
      1   &  \lambda^{r}  &  \lambda^{\psi_1-\psi_2}\\
      \lambda^{r} & 1   &  \lambda^{\psi_1-r-\psi_2} \\
      \lambda^{\psi_1-\psi_2} & \lambda^{\psi_1-r-\psi_2} & 1 
\end{array}
\Rs\sim
\Ls
\begin{array}{ccc}
      1   &  \lambda^{r}  &  \lambda\\
      \lambda^{r} & 1   &  \lambda^{1-r} \\
      \lambda & \lambda^{1-r} & 1 
\end{array}
\Rs.
\label{e6MNS}
\eequ
Because the mass of the light neutrinos is given from the operators, 
 $L_iL_jH_u^2$, we can estimate neutrino mass matrix as
\bequ
M_\nu\sim\left(
\begin{array}{ccc}
      \lambda^{2(\psi_1-\psi_2)} 
   &  \lambda^{2\psi_1-r-2\psi_2}
   &  \lambda^{\psi_1-\psi_2} \\
      \lambda^{2\psi_1-r-2\psi_2} &  \lambda^{2(\psi_1-r-\psi_2)} 
   &  \lambda^{\psi_1-r-\psi_2} \\
      \lambda^{\psi_1-\psi_2} & \lambda^{\psi_1-r-\psi_2} & 1 
\end{array}
\Rs\lambda^{2(\psi_2+\phi-r+\Delta\phi)}
   \frac{\eta^2\VEV{H_u^2}}{\Lambda}, 
\label{e6neutrino}
\eequ
where $\tilde \psi_2({\bf 16,\bar 5})+\tilde \phi({\bf 10,5})
=\psi_2+\phi-r+\Delta\phi$.
Let us introduce a parameter $l$ that parameterize the mass of 
 the heaviest light neutrino, $m_{\nu_3}$, as 
 $-(l+5)\equiv2\psi_2+2(\phi-r+\Delta\phi)$, namely
\bequ
 \lambda^{(l+5)}\sim \frac{\eta^2\VEV{H_u^2}}{m_{\nu_3}\Lambda}.
\eequ
In order to explain the atmospheric neutrino experiments \cite{atmos}, 
 $l$ should be around $-1$-$-4$.

Finally, let us comment on the gauge coupling unification.
In Ref.\cite{NGCU}, it is shown that the success of the gauge coupling 
unification 
 in the minimal SUSY $SU(5)$ model can completely be reproduced 
in the scenario of GUT with an anomalous $U(1)_A$ gauge symmetry 
if the VEV relation (\ref{VEV}) holds, although the mass spectrum of
superheavy fields does not respect the $SU(5)$ gauge symmetry.
In the latter scenario, the effective mass of the color triplet partners of 
 the MSSM doublet Higgs should be around the usual GUT scale 
 $\Lambda_G\sim2\times10^{16}\GeV$, if the contribution from $\order1$ 
 coefficients that we have omitted is negletcted. 
This condition can be expressed as 
\begin{equation}
m_C^\eff\sim\lambda^{2\phi+\Delta\phi}\sim1.
\end{equation}
Because the cutoff scale in the scenario is around the usual GUT scale,
$\Lambda_G$, this condition may lead to too rapid proton decay via
dimension 5 operators. However, in the scenario, the masses of superheavy
fields are determined by the anomalous $U(1)_A$ charges but still have the
ambiguity of the $\order1$ coefficients. Because the number of the superheavy
fields is large, the contribution of the $\order1$ coefficients cannot be
neglected. 
Actually, the ambiguity due to the $\order1$ coefficients makes 
 larger $m_C^\eff$ possible, and indeed we push $m_C^\eff$ 
 larger to suppress the proton decay via dimension 5 operators.
But larger $m_C^\eff$ requires a larger ambiguity which 
 seems less natural.
Thus, as long as the sufficient suppression of the proton decay
is achieved, 
a smaller $m_C^\eff$, which is realized 
 by larger $\phi$ and larger $\Delta\phi$, is preferred for the 
 gauge coupling unification.

\subsection{$E_6\times SU(2)_H$}
\label{e6su2}

Since the matter sector can be unified in terms of the 
 $SU(2)_H$ horizontal symmetry as in \S\ref{horizreview}, 
 the next task is to unify the Higgs sector.
Motivated by the decomposition
 $E_8\supset E_6\times SU(3)_H \supset E_6\times SU(2)_H$, 
 under which {\bf248} of $E_8$ is decomposed as 
\beqn
 {\bf248} &\rightarrow&
   ({\bf78},{\bf1})+({\bf1},{\bf8})
  +({\bf27},{\bf3}) + ({\bf{\cc{27}}},{\bf\bar{3}}) \\
 &\rightarrow&
   ({\bf78},{\bf1})+({\bf1},{\bf3}+{\bf2}+{\bf2}+{\bf1})
  +({\bf27},{\bf2}+{\bf1}) + ({\bf{\cc{27}}},{\bf2}+{\bf1}), 
\eeqn
we assign non-trivial representation of the horizontal symmetry 
 only to {\bf27}, ${\bf\cc{27}}$ and {\bf1} Higgs.
By the way, in the matter sector, two {\bf27} for the 1st and 2nd 
 generations are treated as a doublet, and the difference of 
 their effective charges should correspond 
 to the Cabibbo angle, 
 $\s\psi_1-\s\psi_2\sim1$. 
This means the difference of the effective charges of two 
 components of doublets should be also around 1 
 if the effective charge is well-defined%
\footnote{
Later, we consider models where the effective charge is 
 not well-defined.
But as in the analysis, if the ill-definedness is small,
then  the following discussion can be applied.
}.
In the Table \ref{HiggsContentreduced}, 
 we introduce two {\bf27} ($\Phi, C'$)
 and two ${\bf\cc{27}}$ ($\bar C, \bar C'$), where 
 the primed fields have positive charges and 
 the unprimed fields have negative charges. 
And the difference of anomalous $U(1)_A$ charges of each 
 two fields is much larger than 1. 
Thus, it is difficult to unify the Higgs sector of 
 the Table \ref{HiggsContentreduced}
\footnote{
If we take $\tilde \psi_1-\tilde \psi_2\gg1$, which corresponds 
 $\lambda>\sin\theta_C$, we may obtain 
a suitable model which has realistic quark and lepton masses and
 mixings as in the previous $SO(10)$ case.
 However, it must be difficult for the horizontal symmetry to suppress
 the FCNC processes sufficiently. 
Of course, if the universality of sfermion masses are guaranteed by some 
other mechanism, such models can be realistic. But we do not examine
such possibility here.
}.
And thus, we concentrate on the Higgs sector of the Table \ref{HiggsContentE6}, 
 which contain
\bequ
\begin{array}{cccc}
  {\bf 78} &:&   A,  & A'      \\
  {\bf 27} &:&   \Phi,\ C, &  C'  \\
  {\bf\cc{27}} &:& \bar \Phi,\  \bar C, & \bar C'
\end{array}
\eequ
and singlets.
By the same reason as mentioned in the previous footnote, it is difficult to 
embed 
 the primed fields into a doublet if we aim to suppress 
 the FCNC processes 
 not assuming the universal soft mass.
If we take $(\Phi,C)$ as a doublet under $SU(2)_H$, the Yukawa interaction
for the top quark, 
$\Psi_3\Psi_3\Phi$, is forbidden by the horizontal symmetry.
Thus the remaining possibility is to embed $\bar \Phi$ 
 and $\bar C$ into a doublet as 
 $\bar C_a =(\bar\Phi,\bar C)$.

The Higgs content we consider below is summarized 
 in Table \ref{HiggsContentE6SU2}.
\begin{table}[th]
\caption{
The Higgs content of $E_6\times SU(2)_H\times U(1)_A$ models 
 except for singlets:
Here $SU(2)_H$ doublets are denoted by the index $a$.
All the non-vanishing VEVs are shown, and their magnitudes 
 are formally written by introducing parameters $\Delta\phi$ etc..
One or more discreate symmetries are introduced according to need.
}
\label{HiggsContentE6SU2}
\[
\begin{array}{|c|c|c|} 
\hline
           &   \mbox{non-vanishing VEV}  & \mbox{vanishing VEV} \\
\hline 
  {\bf 78} &   A\Ls \VEV{{\bf45}_A}\sim\lambda^{-a}\Rs  & A'      \\
  {\bf 27} &   \Phi\Ls \VEV{{\bf1}_\Phi}
                   \sim\lambda^{-(\phi-\Delta\phi)}\Rs,\ 
               C\Ls \VEV{{\bf16}_C}\sim\lambda^{-(c-\Delta c)}\Rs 
           &  C'  \\
  {\bf\cc{27}} & \bar C_a\Ls 
                    \VEV{{\bf{1}}_{\bar C_1}}
               \sim \lambda^{-(\bar c+\Delta\phi+\Delta f_c)},\ 
                    \VEV{{\bf\cc{16}}_{\bar C_2}}
               \sim \lambda^{-(\bar c+\Delta c-\Delta\bar f_c)} \Rs 
               & \bar C'\\
  {\bf1} & \bar F_a\Ls \VEV{\bar F_1}
                    \sim\lambda^{-(\bar f+\Delta\bar f)}\Rs,\ 
           F_a \Ls \VEV{F_2}\sim\lambda^{-(f-\Delta f)}\Rs& \\
\hline
\end{array}
\]
\end{table}
For simplicity, we assume that any component fields other than 
 shown in Table \ref{HiggsContentE6SU2} have vanishing VEVs.
If three GUT and horizontal gauge singlets have non-vanishing VEVs as in 
the VEV relations (\ref{VEV}) from three $F$-flatness conditions, for 
example, 
\beqn
  \Phi \bar C F &\sim& \lambda^{-(\phi+\bar c+f)} 
                   \ \equiv \lambda^{-3k},
\label{Fflat1}\\
  C \bar C\bar F &\sim& \lambda^{-(c+\bar c+\bar f)}, 
\label{Fflat2}\\
  F\bar F &\sim& \lambda^{-(f+\bar f)},
\label{Fflat3}
\eeqn
relations $\Delta f=\Delta\bar f=\Delta f_c=\Delta\bar f_c$
are obtained. Once these relations are fixed by three $F$-flatness 
conditions, the VEVs of the other singlet operators that have non-vanishing 
VEVs automatically satisfy the VEV relations (\ref{VEV}).
When some of the non-vanishing VEVs do not satisfy the 
VEV relations (\ref{VEV}), generally, these $\Delta f$s have
different values. Such models will be discussed later, in which
the  effective charge can not be well-defined. 
But for the moments, let us assume 
 Eqs.(\ref{Fflat1})-(\ref{Fflat3}) hold.
In addition to the three relations, 
 three D-flatness conditions
\beqn
  \abs{{\bf1}_\phi}^2 &=& \abs{{\bf1}_{\bar C_1}}^2, \\
  \abs{{\bf16}_C}^2 &=& \abs{{\bf\cc{16}}_{\bar C_2}}^2, \\
  \abs{{\bf1}_{\bar C_1}}^2 + \abs{\bar F_1}^2
 &=& \abs{{\bf\cc{16}}_{\bar C_2}}^2 + \abs{F_2}^2
\label{DcondiSU2}
\eeqn
determine three parameters, $\Delta\phi$, $\Delta c$ and $\Delta f$,
in terms of the anomalous $U(1)$ charges.
Roughly, there are four possible cases as follows:
\begin{enumerate}
 \item ${\bf1}_{\bar C_1} \sim {\bf\cc{16}}_{\bar C_2}
                          \geq \bar F_1,\,F_2 $
 \item $\bar F_1 \sim {\bf\cc{16}}_{\bar C_2} 
           \geq {\bf1}_{\bar C_1},\,F_2 $
 \item $\bar F_1 \sim F_2 
           \geq {\bf1}_{\bar C_1},\,{\bf\cc{16}}_{\bar C_2} $
 \item ${\bf1}_{\bar C_1} \sim F_2 
                         \geq \bar F_1,\,{\bf\cc{16}}_{\bar C_2} $
\label{VacuumStructure}
\end{enumerate}
As for the 2nd and 3rd case, the horizontal breaking scale is 
larger than
the GUT breaking scale $\VEV{\bf 1}$. 
As discussed in \S\ref{e6review},
$\VEV{\bf 1}$ does not seem so small as $\lambda^2$. Therefore, 
 $SU(2)_H$ breaking scale larger than $\VEV{\bf1}$ is not sufficient
 for the suppression of the FCNC processes.
For simplicity,  we concentrate on the 4th case in the following discussion,
but a similar discussion can be applied to the other cases.
In the 4th case,
\beqn
  {\bf1}_\Phi = {\bf1}_{\bar C_1} \sim F_2 &\sim& \lambda^{-k} ,\\
  \bar F_1 &\sim& \lambda^{-(f+\bar f)+k} , \\
  {\bf16}_C = {\bf\cc{16}}_{\bar C_2} 
            &\sim& \lambda^{\half[-(c+\bar c)+f-k]}, 
\eeqn
in other words,
\beqn
 \Delta f &=& \frac{2f-\phi-\bar c}{3} = f-k ,
\label{deltaf}\\
 \Delta \phi &=& \frac{2\phi -f-\bar c}{3} = \phi-k ,\\
 \Delta c &=& \frac{c-\bar c +\Delta f}{2}.
\eeqn
The condition for the 4th case ($F_2 
 \geq \bar F_1,\,{\bf\cc{16}}_{\bar C_2} $) to be realized is given by 
\bequ
 0<-k \leq -\half(f+\bar f)\,,\,\, -c-\bar c+f,
\label{4thCase}
\eequ
which is also written as 
\bequ
 f< \Delta f \leq \half(f-\bar f)\,,\,\, -c-\bar c+2f.
\label{4thCase'}
\eequ

In addition, as shown in \S\ref{e6review}, the following 
 conditions are required phenomenologically:
\bit
 \item The parameter $r$ for the neutrino mixings should be 
       around $\half$-$\frac{3}{2}$. 
 \item The parameter $l$ for the neutrino mass scale should be 
       around $-1$-$-4$.
 \item In order to realize the DT splitting, 
       $C'A\Phi\Phi$ must be allowed, 
       and $C'AC\Phi$ must be forbidden.
 \item $\bar C F\bar C F\bar C\bar F$, which corresponds to 
       $\bar\Phi\bar\Phi\bar C$ in \S\ref{e6review}, must be 
       allowed in order to avoid undesired massless modes.
 \item In order to give mass to would-be PNG modes, 
       $A'A\Phi\bar C F$ must be allowed. 
 \item For the gauge coupling unification, smaller 
       effective mass of the colored Higgs, 
       $m^\eff_C\sim\lambda^{2\phi+\Delta\phi}$, is preferable.
       In the model displayed in Table \ref{HiggsContentE6}, 
       $m^\eff_C\sim\lambda^{-8.5}$.
 \item In order to reproduce the realistic quark mass matrices,
       the $SU(2)_R$ symmetry must be broken in the Yukawa couplings.
       $SU(2)_R$ breaking VEVs $\VEV{C}=\VEV{\bar C}$ can be 
       picked up through the SM Higgs mixing 
       ($\bar C'\bar C\bar FA\Phi^2$ is required), 
       or through higher dimensional interactions 
       (for example, $\Psi\bar CC\Psi\bar F\Phi$)
\footnote{
Although there are $SU(2)_R$ breaking effects in the 
 $3\times6$ mass matrix of ${\bf5}$-${\bf{\bar5}}$ components 
 of $\Psi_a$ and $\Psi_3$, 
 it is not sufficient to reproduce the realistic quark mixings%
\cite{E6}.
Thus we need another source of $SU(2)_R$ breaking.
}.       
\eit
These conditions are rewritten 
 in terms of the anomalous $U(1)$ charges as
\begin{eqnarray}
&&\half
         \lesssim r=\half(c-\phi)+\Delta f
         \lesssim \frac{3}{2} 
\label{1st}
\\
&&-1\gtrsim l=-5-2(\psi-\Delta f-\psi_3)+\phi+2\Delta c
                   \gtrsim -4
\label{2nd}\\
&&c<\phi \label{3rd}\\
&&\bar f \geq -3\bar c-2f \label{4th}\\
&&0\leq a'+a+\phi+\bar c+f\geq 0 \label{5th}\\
&&2\phi+\Delta\phi\gtrsim -8.5 \label{6th}\\
&&2\psi+\phi+c+\bar c+\bar f\geq 0\quad {\rm or}
 \quad \bar c'\geq -2\phi-\bar c-\bar f-a,
\hspace{5cm}
\label{7th}
\end{eqnarray}
Note that the 1st condition is not consistent with the 3rd conditions
if $\tilde \psi_1-\tilde \psi_2=1$, that is, $\Delta f=\half$ to reproduce
the suitable value of the Cabibbo angle.
There are three ways to avoid this inconsistency\footnote{
As in the previous $SO(10)$ model, we can take 
$\tilde\psi_1-\tilde \psi_2>1$ to avoid this inconsistency. Because
the atmospheric neutrino mixings tend to be rather small, here, we 
do not examine this possibility.}.
\begin{enumerate}
 \item To relax the 1st requirement. \\
       For example, $r=\frac{1}{4}$ is not an unacceptable choice, 
       although rather large ambiguity of 
       $\order1$ coefficients are needed
       to reproduce the large atmospheric neutrino oscillation.
 \item To set $c\geq\phi$ and introduce an additional discrete symmetry 
       to forbid $C'AC\Phi$.\\
       In this case, $r=\half(c-\phi+2\Delta f)$.
       If $c=\phi$ is taken, the relation 
       $r=\half$ is obtained.
 \item To give up the effective charge.\\
       In terms of the notation 
       in Table \ref{HiggsContentE6SU2}, $r$ is given by
\[
 r=\Delta c-\Delta\phi=\frac{c-\phi+\Delta f_c+\Delta\bar f_c}{2}.
\]
       Thus, if $\Delta f_c+\Delta\bar f_c > \psi_1-\psi_2$, 
       sufficiently large $r$ can be obtained without imposing 
       additional symmetries.
\end{enumerate}
Along each strategy, we construct realistic models as follows.

\subsubsection{$c<\phi$ ($r<\half$)}
\label{strategy1}
In this analysis, we fix the $\Delta f$ as $\half$. 
The relation $r=\half(c-\phi+1)$ indicates that 
the larger $c-\phi<0$ leads to
larger $r<\half$. Therefore, if $c-\phi$ is taken 
as the minus minimum unit of 
$U(1)_A$ charge, then $r$ becomes the closest value to $\half$.
Therefore, the smaller unit leads to the closer value 
of $r$ to $\half$.
Here, we introduce half integer $U(1)_A$ charges 
and take $\theta=-\half$, 
which give $r=\frac{1}{4}$.

As noted before, in the vacuum 4, the $SU(2)_H$ breaking scale is the
same as the $E_6$ breaking scale, because the VEVs 
$\VEV{{\bf 1}_\Phi}=\VEV{{\bf1}_{\bar C_1}}
               \sim \VEV{F_2}\sim \lambda^{-k}$ 
break simultaneously $SU(2)_H$ and $E_6$. In order to suppress 
the FCNC processes, a smaller $SU(2)_H$ breaking scale  is preferable, 
but on the other hand,
a smaller $E_6$ breaking scale leads to 
a larger effective colored Higgs mass,
which can spoil the success of the gauge coupling unification 
and/or result in
the non-perturbative region of gauge couplings, as noted 
in \S\ref{e6review}.
Taking account of the above conflict,  we take $k=-1$ here. 
Thus, the relation $k=f-\Delta f$ leads to $f=-\half$.

Then, the condition for the vacuum structure \ref{VacuumStructure}, 
 Eq.(\ref{4thCase}), and the condition (\ref{4th}) give a relation 
\bequ
 2k-f \geq -3\bar c-2f, 
\label{cbar}
\eequ
that is, $\bar c\geq\frac{5}{6}$.
Under fixed $k$ and $f$, because $3k=\bar c+f+\phi$, 
larger $\bar c$ leads to smaller $\phi$,
and therefore a larger colored Higgs mass, which leads to less natural
explanation for the success of the gauge coupling unification. 
Therefore, we adopt 
$\bar c=1$, which leads to 
 $\phi=-\frac{7}{2}$ and $c=-4$.

Now, Eq.(\ref{4thCase}) and $\bar f \geq -3\bar c-2f$ lead to 
 $-2\leq \bar f\leq-\frac{3}{2}$.
And we take $\bar f=-2$.

As for $a$, $a=-\half$ and $a=-1$ are possible.
The former yields relatively large FCNC processes 
because $\VEV{A}$ breaks the $SU(2)_E$ symmetry 
 which guarantees the universality of masses of three 
 ${\bf \bar 5}$ sfermion fields. 
Therefore, we take $a=-1$, though the gauge couplings may become in 
non-perturbative region.

Table \ref{ContentStrategy12} shows examples for the latter case.
\begin{table}
\caption{
Examples of the charge assignments for the 1st and 2nd strategies : 
Signs denote the additional $Z_2$ symmetry 
 that play the same role as the $Z_2$ symmetry 
 introduced in Table \ref{HiggsContentE6}. 
This charge assignment yields $r=\half+\frac{c-\phi}{2}$ and 
 $l=-5-c$.
Odd quarter integer charges of the matter fields ($\Psi_3$, $\Psi_a$) 
 guarantees that the $R$-parity is automatically conserved.
When $c\geq\phi$, we impose an additional $Z_2$ symmetry 
 and introduce a singlet field $Z_C$ 
 to forbit $C'AC\Phi$ while allowing $C'A\Phi\Phi$. 
}
\label{ContentStrategy12}
\[
\begin{array}{|c|c|c|} 
\hline
                  &   \mbox{non-vanishing VEV}  
                  &   \mbox{vanishing VEV} \\
\hline 
{\bf 78}          &   A(a=-1;-)        & A'(a'=5;-)  \\
{\bf 27}          &   \Phi(\phi=-7/2;+),\,  C(c=-4,-7/2,-3,-5/2;+) 
                  &   C'(c'=8;-)  \\
                 &&   \Psi_3 (\psi_3=7/4;+),\,\Psi_a (\psi=17/4;+)\\
{\bf \cc{27}}     &   \bar C_a(\bar c=1;+) 
                  &   \bar C'(\bar c'=11/2;-) \\
{\bf 1}           &   \bar F_a (\bar f=-2;+),\,F_a(f=-1/2;+) &\\
                  &   \Theta (\theta=-1/2;+),\, Z_i(z_i=-3/2;-) &  \\
                  &   Z_C(z_C=\mbox{-},-1/2,-1,-3/2;+) &  \\
\hline
\end{array}
\]
\end{table}
$(a',c',\bar c')$ are determined by the smallest values that 
 allow $A'A^5$, $A'\Phi\bar CF$, $C'A\Phi\Phi$ and 
 $\bar C'ZC$.
We set $z$ the largest value which forbids $C'Z\Phi\Phi$.
Here, the matter fields ($\Psi_3$, $\Psi_a$) are also shown.
From their charges, we can find $l=-1$ and 
 $\Psi\Psi\Phi C\bar C\bar F$ is allowed, which introduces $SU(2)_R$ 
 breaking in the Yukawa couplings.
Note that only the matter fields have odd quarter integer charges, 
 and therefore they always appear in pair, 
 which guarantees the $R$-parity is automatically conserved.
The effective colored Higgs mass is given 
 as $\lambda^{-19/2}$. This value is not so much different from the 
 value of the model in Table \ref{HiggsContentE6}.
And the parameter $\delta_{\bf10}$ and $\delta_{\bf{\bar5}}$ 
 are estimated as 
\bequ
\delta_{\bf10} \sim 
 \begin{pmatrix}
   \lambda^2 & \lambda^3 & \lambda^3 \\
   \lambda^3 & \lambda^2 & \lambda^2 \\
   \lambda^3 & \lambda^2 & 1 
 \end{pmatrix}\,,\quad
\delta_{\bf{\bar5}} \sim 
 \begin{pmatrix}
   \lambda^2     & \lambda^{2+r} & \lambda^3 \\
   \lambda^{2+r} & \lambda^2     & \lambda^{3-r} \\
   \lambda^3     & \lambda^{3-r} & \lambda^2 
 \end{pmatrix},
\label{e6su2delta}
\eequ
which are obtained from the non-renormalizable interactions, for example,
\begin{equation}
\int d^4\theta Z^\dagger Z[|\Psi F|^2+|\Psi F^\dagger|^2
+\Psi^\dagger A^2\Psi].
\end{equation}
Off diagonal elements of $\delta$ become smaller than those in
the previous $SO(10)$ model, though these are still too large to suppress
the FCNC processes and we must require other mechanisms that 
suppresses the above non-renormalizable interactions with the spurion 
field $Z$ or that gives universal sfermion masses.

\subsubsection{$\phi\leq c$ $(r\geq \half)$}
\label{strategy2}

Next, let us examine the 2nd strategy.
With the aid of an additional discrete symmetry, we can 
 forbit the interaction $C'AC\Phi$ while allowing $C'A\Phi\Phi$ 
 even when $\phi\leq c$, which always leads to $r\geq\half$.
For example, consider another $Z_2$ symmetry that only $C$ and 
$Z_C$ have odd parity. Here $z_C<\phi-c$ is required to forbid
$C'AC\Phi Z_C$ and to allow $C'A\Phi\Phi$.

In this analysis, we also introduce half-integer charges and 
 fix the value of $\Delta f$ as $\half$.
Then, as in the previous strategy, we set 
 $(k, f, \bar c,\phi,\bar f,a)=(-1,-\half,1,-\frac{7}{2},-2,-1)$.
For these charges, Eq.(\ref{4thCase}) requires 
 $c\leq k-\bar c+f=-\frac{5}{2}$, 
 which leads to $c=-\frac{7}{2},-3,-\frac52$.
$(a',c',z)$ are also determined as in the previous 
 strategy.
We set $z_C$ as the largest negative value satisfying $\phi>c+z_C$, and 
 $\bar c'$ is determined to allow $\bar C'ZCZ_C$%
\footnote{
Another choice is to assign odd parity to $\bar C$ and 
 determine $\bar c'$ so as to $\bar C'ZC$  allowed. 
This choice is convenient for embedding the model into 
 $E_6\times SU(3)_H$ model, 
 and we consider this possibility later.
}.
Table \ref{ContentStrategy12} summarizes the charge assignments.

Here, the matter fields ($\Psi_3$, $\Psi_a$) are also shown.
From their charges, 
 we can find $l=-3/2,-2,-5/2$ and that 
 $\Psi\Psi\Phi CZ_C\bar C\bar F$ is allowed, which is
 important to introduce $SU(2)_R$ breaking in Yukawa couplings.
Again, the $R$-parity is automatically conserved.
The effective colored Higgs mass is given 
 as $\lambda^{-19/2}$.
And the parameter $\delta_{\bf10}$ and $\delta_{\bf{\bar5}}$ 
 are given by the same expression as in Eqs.(\ref{e6su2delta}).

\subsubsection{$\Delta f\neq\Delta\bar f$}
\label{strategy}
Finally, we investigate the possibility where 
 the effective charge is not well-defined.
It means the VEV relation (\ref{VEV}) 
does not hold generally.
Although this relation is naturally expected to hold, 
 the ambiguity of $\order1$ coefficients may shift 
 the relation slightly.
Here, we assume one of the three singlet operators in Eqs.(\ref{Fflat1})-
(\ref{Fflat3}) has
smaller non-vanishing VEV than given in the equaitons
because of a cancellation.

From the definition 
$\lambda^r\equiv\frac{\lambda^\phi\VEV{\Phi}}{\lambda^c\VEV{C}}$, 
we can see the smaller $\VEV{{\bf16}}$ 
 and/or the larger $\VEV{{\bf1}}$ result in larger $r$.
Thus, if the cancellation occurs in Eq.(\ref{Fflat1}), 
 smaller $r$ will be resulted.
If the cancellation occurs in Eq.(\ref{Fflat2}), 
 $r$ will be larger. However, in this case, $SU(2)_R$ breaking
 scale becomes smaller than expected by the charges, that is,
 the $SU(2)_R$ breaking effects in Yukawa couplings become
 smaller, which results in unrealistic quark mass matrices.
Therefore, we assume
\bequ
 F\bar F\,\sim\,\epsilon\lambda^{-(f+\bar f)},
\eequ
where $\epsilon\ll1$, instead of Eq.(\ref{Fflat3}).
Eqs.(\ref{Fflat1}) and (\ref{Fflat2}) mean
 $\Delta f=\Delta f_c$ and $\Delta\bar f=\Delta\bar f_c$, 
 and therefore $\epsilon\sim\lambda^{\Delta f-\Delta\bar f}
                        \equiv \lambda^\delta$. 
$\epsilon \ll 1$ leads to $\Delta f > \Delta\bar f$.
In this case, magnitudes of couplings in the low energy effective theory 
 depend on which interactions exist, in contrast to the case 
 where the effective charge is well-defined and the magnitudes 
 are written by a simple sum of the effective charge. 
Let us illustrate this by using the Yukawa matrix, $\Psi\Psi\Phi$
 as a concrete example.
For simplicity, we set $\psi_3=n=-\phi/2$ and $\psi=n+m$.
Then the Yukawa matrix in terms of exponent becomes
\bequ
\Ls
\begin{array}{ccc}
 2m+\Lm
  \begin{array}{l}
  2\Delta f\\ \Delta f+\Delta\bar f \\ 2\Delta\bar f
  \end{array}
 \RR
&
 2m+\Lm
  \begin{array}{l}
  \Delta f-\Delta\bar f\\ 0 \\-\Delta f+\Delta\bar f
  \end{array}
 \RR
&
 m+\Lm
  \begin{array}{l}
  \Delta f\\ \Delta\bar f 
  \end{array}
 \RR
\\
 2m+\Lm
  \begin{array}{l}
  \Delta f-\Delta\bar f\\ 0 \\-\Delta f+\Delta\bar f
  \end{array}
 \RR
&
 2m+\Lm
  \begin{array}{l}
  -2\Delta\bar f\\ -\Delta f-\Delta\bar f \\ -2\Delta f
  \end{array}
 \RR
&
 m+\Lm
  \begin{array}{l}
  -\Delta\bar f\\ -\Delta f 
  \end{array}
 \RR
\\
 m+\Lm
  \begin{array}{l}
  \Delta f\\ \Delta\bar f 
  \end{array}
 \RR
&
 m+\Lm
  \begin{array}{l}
  -\Delta\bar f\\ -\Delta f 
  \end{array}
 \RR
&
0
\end{array}
\Rs,
\label{YukawaMatrix}
\eequ
where the term, for example, $m-\Delta\bar f$ denotes the element of 
 Yukawa matrix is order of $\lambda^{m-\Delta\bar f}$.
Here, the exponents $-\Delta\bar f$, $+\Delta f$, $-\Delta f$ and 
$+\Delta\bar f$ come from the VEVs 
$\VEV{\bar F}\sim \lambda^{-\bar f-\Delta\bar f}$, 
$\VEV{F}\sim \lambda^{-f+\Delta f}$, 
$\VEV{\Phi\bar C}\sim \lambda^{-\phi-\bar c-\Delta f}$ and 
$\VEV{C\bar C}\sim\lambda^{-\bar c-c+\Delta\bar f}$, respectively.
One of the terms of each element of 
the matrix (\ref{YukawaMatrix}) is realized.
This depends on which operator are allowed by the symmetry.
For example, for the (1,1) element for the Yukawa matrix, interactions 
$\lambda^{2\psi+2f+\phi}\Psi F\Psi F\Phi$, 
$\lambda^{2\psi+f+\bar c+2\phi}\Psi F\Psi \bar C\Phi\Phi$
and $\lambda^{2\psi+2\bar c+3\phi}\Psi\bar C\Phi\Psi\bar C\Phi\Phi$, which
are allowed if 
$2\psi+2f+\phi\geq 0$, 
$2\psi+f+\bar c+2\phi\geq 0$
and $2\psi+2\bar c+3\phi\geq 0$ are satisfied, induce
$\lambda^{2m+2\Delta f}$, 
$\lambda^{2m+\Delta\bar f+\Delta f}$
and $\lambda^{2m+2\Delta\bar f}$, respectively.
That is, the terms that include $\Phi\bar C$ ($C\bar C$) have
$\lambda^{-\delta}$ times larger couplings 
than the terms that include 
$\bar F$ ($F$) instead of $\Phi\bar C$ ($C\bar C$).

In the following arguments, we assume $\Delta\bar f=\half$ and we discuss a 
possibility that $\lambda^{2\Delta\bar f}\sim 0.22$ becomes 
the Cabibbo angle\footnote{
In order to obtain $r\geq \half$, $\Delta f+\Delta\bar f>1$ is required
if $c-\phi<0$. In that situation, to obtain the suitable value for the
Cabibbo mixing, it is a reasonable requirement that either of $\Delta f$ and
$\Delta\bar f$ becomes $\half$. Because $\Delta f>\Delta\bar f$, 
we assume that $\Delta\bar f=\half$ here.}.
Here we introduce half integer charges and set $c=\phi-\half$. 
In order to yield the suitable Cabibbo angle, the difference 
 between the (2,2) element and the (1,2) element should be 
 $2\Delta\bar f$, 
 and thus the (1,2) element is $2m$ or $2m-\Delta f+\Delta\bar f$,
 and the (2,2) element is $2m-2\Delta\bar f$ or $2m-\Delta\bar f-\Delta f$,
 respectively. 
If $\phi+\bar c>\bar f$, when $\Psi\Psi\Phi\VEV{\bar F^2}$ or 
$\Psi\Psi\Phi\VEV{\Phi\bar C\bar F}$ is allowed by the symmetry,
then $\Psi\Psi\Phi\VEV{\Phi^2\bar C^2}$ is also allowed, which results in
$2m-2\Delta f$ in the (2,2) element. Therefore, we assume that
\begin{equation}
\phi+\bar c<\bar f.
\label{phicf}
\end{equation}
The latter of the (1,2) element comes from 
$\Psi\Psi\Phi\VEV{\Phi\bar CC\bar C}$, which is allowd if
 $2m\geq-(\phi+c+2\bar c)=-2(\phi+\bar c)+\half$. 
Then, $\Psi\Psi\Phi\VEV{\Phi\bar C\Phi\bar C}$ 
 is also allowed, which results in 
 $2m-2\Delta f$ in the (2,2) element. 
And thus, the Cabibbo angle becomes 
 too small. Therefore, the (2,2) and (1,2) elements must be
 $2m-2\Delta\bar f$ and $2m$, respectively.
And if the (2,3) element is $m-\Delta f$, which comes from 
 $\Psi_3\Psi\Phi\VEV{\Phi\bar C}$, a charge relation, $m\geq-(\phi+\bar c)$, 
must be satisfied, and thus $\Psi\Psi\Phi\VEV{\Phi\bar C\Phi\bar C}$ 
 is also allowed, which must be forbidden to reproduce the suitable
 Cabibbo angle.
In summary,  
 the following elements are selected : 
\bequ
\Ls
\begin{array}{ccc}
\mbox{``any''} & 2m & \mbox{``any''} \\
2m & 2m-2\Delta\bar f & m-\Delta\bar f \\
\mbox{``any''} &m-\Delta\bar f & 0
\end{array}
\Rs, 
\eequ
where ``any'' suggests any terms are possible. 
This condition is expressed in terms of the anomalous $U(1)$ 
 charges as 
\begin{equation}
\phi+\bar c+\bar f<-2m
\label{8th}
\end{equation}
to forbit the term $\Psi\Psi\Phi\VEV{\Phi\bar C\bar F}$ which yealds 
       $2m-\Delta{f}-\Delta\bar f$ for the (2,2) element.
Together with the condition (\ref{4th}) 
 and the definition of $k$, $3k=\phi+\bar c+f$, 
 we get 
\bequ
 \phi < 2k-\frac{2}{3}m.
\label{rightCabibbo}
\eequ
By the way, when $\phi+\bar c+\bar f<-2m$, $\phi>c$, 
 and Eq. (\ref{phicf}) hold, 
 $m$ is smaller than $-(c+\bar c)$, that is 
$\Psi_3\Psi\Phi\VEV{\bar CC}$ is forbidden by the SUSY-zero mechanism. 
This means that the (1,3) component of the matrix 
 becomes $m+\Delta f$ that is induced from the interaction 
 $\Psi\Psi_3\Phi F$, or the (1,3) element of the corresponding Yukawa
 matrix vanishes.
The important point is that the (1,3) element is larger than the expected
value, $(2,3)+(1,2)-(2,2)=m+\Delta\bar f$, that is, the (1,3) element of the
corresponding Yukawa matrix is smaller than the expected value.
Then without the term $\Psi_3\Psi C\bar F$, 
 the down quark mass and the electron
mass become too small (see Appendix \ref{55barmass} for the detail arguments).
Therefore, the term $\Psi_3\Psi C\bar F$ is required. 
This requirement is written as 
\begin{equation}
 \bar f\geq-(m+c-\phi)=-m+\half.
\end{equation}
Together with the above condition (\ref{8th}),
 we get
\bequ
 \bar c < -\phi-m-\half.
\label{cbarStrategy3}
\eequ

Here, we impose one more condition that keep the gauge couplings 
 in the perturbative region.
This is realized by $a=-\half$.
Then, in order to allow $A'A\Phi\bar C F$ to give mass to 
 PNG modes while forbidding $A'A^7$, 
 we should take $k\geq-5/6$.
Because smaller $k$ is preferred in order to suppress the FCNC processes, 
 we set the smallest possible value $k=-\frac{5}{6}$.

Because $m$ should be around $\frac{5}{2}$-$3$ to obtain the hierarchy between
the up-type quark masses of the 2nd and 3rd generation, 
 the condition (\ref{rightCabibbo}) leads to 
 $\phi\lesssim-\frac{10}{3}$, and we set $\phi=-\frac{7}{2}$.
If we take $m=\frac{5}{2}$, 
 $\bar c\leq0$ is led from (\ref{cbarStrategy3}), 
 resulting $f\geq1$ due to the definition of $k$. 
Since a larger $f$ leads to a larger $\Delta f$ 
 and therefore a smaller $\epsilon$, 
 which becomes less natural to realize.
Thus we take $\bar c=0$ and $f=1$.
Then, $\delta=\frac{4}{3}$ and $r=\frac{11}{12}$.
And the condition $\VEV{\bf1}\sim\lambda^{-k}\geq\VEV{\bar F}$ 
 and $\bar f \geq -3\bar c-2f$ make $-4/3\geq \bar f\geq -2$, 
 and we fix $\bar f=-2$.

Table \ref{ContentStrategy3} summarizes the charge assignment.
\begin{table}
\caption{
An example of the charge assignments for the 3rd strategy : 
Signs denote the $Z_2$ symmetry 
 that play the same role as the $Z_2$ symmetry 
 introduced in Table \ref{HiggsContentE6}. 
We assume $\VEV{F\bar F}\sim\lambda^\delta\lambda^{-(f+\bar f)}$, 
 $\delta=4/3$ and $\lambda\sim\sin\theta_C$.
This charge assignment yields $r=11/12$ and $l=-2-\delta$.
Odd quarter integer charges of the matter fields ($\Psi_3$, $\Psi_a$) 
 guarantees that the $R$-parity is automatically conserved.
}
\label{ContentStrategy3}
\[
\begin{array}{|c|c|c|} 
\hline
                  &   \mbox{non-vanishing VEV}  
                  &   \mbox{vanishing VEV} \\
\hline 
{\bf 78}          &   A(a=-1/2;-)        & A'(a'=3;-)  \\
{\bf 27}          &   \Phi(\phi=-7/2;+),\,  C(c=-4;+) 
                  &   C'(c'=15/2;-)  \\
                 &&   \Psi_3 (\psi_3=7/4;+),\,\Psi_a (\psi=17/4;+)\\
{\bf \cc{27}}     &   \bar C_a(\bar c=0;+) 
                  &   \bar C'(\bar c'=19/2;-) \\
{\bf 1}           &   \bar F_a (\bar f=-2;+),\,F_a(f=1;+) &\\
                  &   \Theta (\theta=-1/2;+),\, Z_i(z_i=-1;-) &  \\
\hline
\end{array}
\]
\end{table}
$(a',c',z)$ are determined as in the previous strategies in 
\S\ref{strategy1} and \S\ref{strategy2}, while 
 $\bar c'$ is determined to allow $\bar C'\bar C\bar FA\Phi\Phi$.
This term is required to mix ${\bf{16}}_C$ component into 
 the massless doublet Higgs, which brings $SU(2)_R$ breaking 
 effect into the Yukawa couplings.
This $SU(2)_R$ breaking effect is needed because in this case 
 $\Psi\Psi\Phi C\bar C\bar F$ is forbidden.

Note that 
the interactions which have the total charges larger than
$-(c+\phi+2\bar c)$ can couple with a singlet operator
$\Phi C\bar C^2$. 
Because 
$\lambda^{\phi+c+2\bar c}\VEV{\Phi C\bar C^2}\sim \epsilon^{-1}\gg1$,
the coefficients of such interactions are enhanced by factor $\epsilon^{-1}$,
In the Majorana mass matrix and the superheavy Higgs 
mass matrices, 
such interactions appear and the detail analysis 
is done in Appendix \ref{diff}.
The results are that the parameter $l$ shifts from the naively 
 evaluated value by $-\delta$, and the gauge coupling 
 unification is  realized more naturally for the fixed $m^\eff_C$, 
 which is given as $\lambda^{-\frac{37}{4}}$%
\footnote{
In this case, as shown in Appendix \ref{diff}, the main mode
of $H_u$ is included in ${\bf10}_\Phi$ and that of $H_d$ is
in ${\bf16}_C$.
And $m^\eff_C$ is given by the relation $\lambda^{2c+\phi}\VEV C$.
}.
And the parameter $\delta_{\bf10}$ and $\delta_{\bf{\bar5}}$ 
 are estimated as 
\bequ
\delta_{\bf10} \sim 
 \begin{pmatrix}
   \lambda^{5/3} & \lambda^{8/3} & \lambda^3 \\
   \lambda^{8/3} & \lambda^{5/3} & \lambda^2 \\
   \lambda^3     & \lambda^2     & 1 
 \end{pmatrix}\,,\quad
\delta_{\bf{\bar5}} \sim 
 \begin{pmatrix}
   \lambda^{5/3} & \lambda^{1+r} & \lambda^2 \\
   \lambda^{1+r} & \lambda       & \lambda^{2-r} \\
   \lambda^2     & \lambda^{2-r} & \lambda^{5/3} 
 \end{pmatrix}.
\eequ
These values give larger FCNC processes than those obtained by the
1st and the 2nd strategy (\ref{e6su2delta}). This is because we adopt
$a=-\half$, which is required in order to suppress the divergence of the 
gauge couplings. The difference from the previous cases is that
due to the special charge assignment $\bar CC\bar F$ cannot appear 
in the Yukawa
interactions, and therefore, the $SU(2)_R$ breaking must be realized 
through the mixings in the MSSM Higgs. This requirement results in the larger
$U(1)_A$ charge for $\bar C'$ field, which increases the gauge couplings at
the high energy scale.

\subsection{$E_6\times SU(3)_H$}
\label{e6su3}
In this subsection, we consider $E_6\times SU(3)_H$ model, where three $\Psi$ 
 and three ${\bf{\cc{27}}}$ ($\bar C,\bar\Phi,\bar C'$) are 
 a triplet and an anti-triplet of $SU(3)_H$, respectively.
In this case, the anomaly of $SU(3)_H$ of the matter 
 sector is cancelled by that of the three ${\bf{\cc{27}}}$, 
 in contrast to the case of Ref.\cite{horiz} where some additional
fields must be required for the anomaly cancellation.

In order to yield the large top Yukawa coupling, $SU(3)_H$ should be 
 broken near the cutoff scale.
Suppose that $SU(3)_H$ is broken into $SU(2)_H$ at the cutoff 
scale by the VEVs
$\VEV{E}=\VEV{\bar E}\sim \lambda^{-\frac{1}{2}(e+\bar e)}=1$ and 
the effective charges can be defined, that is,
$e+\bar e=0$ is satisfied. 
Then it can be shown that the effective theory with
$SU(2)_H$ can be identified with a certain $SU(2)_H$ model 
that have the same $U(1)_A$ charges as the effective charges 
in the effective $SU(2)_H$ model. 
The essential point is that all the interactions in the $SU(2)_H$
model can be induced from the interactions in the $SU(3)_H$ model.
For example, $\lambda^{2\psi_3+\phi}\Psi_3\Psi_3\Phi$ in $SU(2)_H$ model
can be obtained
from the interaction 
$\lambda^{2\psi+2\bar e+\phi}\Psi \bar E\Psi \bar E\Phi$
by developing the VEV $\VEV{\bar E}\sim 1$.
Note that the coefficient of the effective interaction 
is determined by the effective charges, that is, 
$\lambda^{2\psi+2\bar e+\phi}\VEV{\bar E}^2
 \sim \lambda^{2\tilde \psi_3+\phi}$,
where $\tilde \psi_3$ is the effective charge of $\Psi_3$ of 
the effective $SU(2)_H$ model.
Therefore, 
it is obvious that the total charge of a interaction 
in the $SU(3)_H$ models nothing but 
the total effective charge of the corresponding interaction
in the effective $SU(2)_H$ model because 
$SU(3)_H$ is broken at the cutoff scale. 
Thus, if a term is forbidden 
by the SUSY-zero mechanism in the $SU(3)_H$ model,
the corresponding term in the $SU(2)_H$ model 
is also forbidden by the SUSY-zero mechanism. 
Hence, 
the effective $SU(2)_H$ model can be described
 by the $SU(2)_H$ model.
Conversely, if an $SU(2)_H$ model is found 
in which the $U(1)_A$ charges are 
the same as the effective charges of an $SU(3)_H$ model, 
then an $SU(3)_H$ model can be found straightforwardly. 
Note that for $SU(2)_H$ models, the arguments 
in the previous section can be
applied, which makes the discussion much simpler.

\subsubsection{$SU(2)_H$ models for $SU(3)_H$ models}
\label{Su2ForSU3}
In order to extend the horizontal symmetry to $SU(3)_H$, the difference
$m=\psi-\psi_3$ is required to be the same as $\bar m\equiv \bar c'-\bar c$.
In the 3rd strategy, a large $\bar c'$ is needed to allow 
 $\bar C'\bar C\bar FA\Phi\Phi$, and it is difficult to satisfy the above 
 requirement. 
 Thus we examine the 1st and 2nd strategies in following.

The charge assignments shown 
 in Table \ref{ContentStrategy12} 
 still have discrepancy between  
 $m = \psi-\psi_3 = \frac{5}{2}$ and 
 $\bar m=\bar c'-\bar c = \frac{9}{2}$.
Note that phenomenologically viable value of $m$ is around 
 $\frac{5}{2}$-$3$.
Thus, models with smaller $\bar m$ is needed. 
Since $(f,z_i,z_C,\bar c')$ are set as 
 $(k+\Delta f,a-\half,\phi-c-\half,-c-z_C-z_i)$ in the 1st and 2nd strategies,
  $\bar m$ is written as 
\bequ
 \bar m = \Ls\half-\phi-\Ls a-\half\Rs\Rs-\bar c 
        = 2\times\half-3k+f-a
        = 2\times\half+\Delta f-2k-a. 
\label{mbar}
\eequ
This means that in order to obtain a smaller $\bar m$, 
 larger $a$ and $k$ are required. 
We can construct such models (see Table \ref{ContentSU2forSU3-1}), 
 although the FCNC processes are not suppressed sufficiently:
\bequ
\delta_{\bf10} \sim 
 \begin{pmatrix}
   \lambda & \lambda^{2} & \lambda^3 \\
   \lambda^{2} & \lambda & \lambda^2 \\
   \lambda^3     & \lambda^2     & 1 
 \end{pmatrix}\,,\quad
\delta_{\bf{\bar5}} \sim 
 \begin{pmatrix}
   \lambda & \lambda^{1+r} & \lambda^2 \\
   \lambda^{1+r} & \lambda       & \lambda^{2-r} \\
   \lambda^2     & \lambda^{2-r} & \lambda 
 \end{pmatrix}.
\eequ

\begin{table}[ht]
\caption{
Examples of the charge assignments of $SU(2)_H$ models that 
 can be embedded into $SU(3)_H$ models : 
Signs denote the $Z_2$ symmetry 
 that play the same role as the $Z_2$ symmetry 
 introduced in Table \ref{HiggsContentE6}. 
When $c\geq\phi$, we impose an additional $Z_2$ symmetry 
 and introduce a singlet field $Z_C$. 
These models result in the universal sfermion masses, but the degree
of the universality is not sufficient to suppress the FCNC processes.
}
\label{ContentSU2forSU3-1}
\[
\begin{array}{|c|c|c|} 
\hline
                  &   \mbox{non-vanishing VEV}  
                  &   \mbox{vanishing VEV} \\
\hline 
{\bf 78}          &   A(a=-1/2;-)        & A'(a'=3;-)  \\
{\bf 27}          &   \Phi(\phi=-7/2;+),\,  C(c=-4,-7/2;+) 
                  &   C'(c'=15/2;-)  \\
                 &&   \Psi_3 (\psi_3=7/4;+),\,\Psi_a (\psi=19/4;+)\\
{\bf \cc{27}}     &   \bar C_a(\bar c=2;+) 
                  &   \bar C'(\bar c'=5;-) \\
{\bf 1}           &   \bar F_a (\bar f=-2;+),\,F_a(f=0;+) &\\
                  &   \Theta (\theta=-1/2;+),\, Z_i(z_i=-1;-) &  \\
                  &   Z_C(z_C=\mbox{-},-1/2;+) &  \\
\hline
\end{array}
\]
\end{table}

In order to improve the suppression of the FCNC processes, we have to 
change some assumptions.
If we employ the other choice of $Z_2$ parity introduced in \S\ref{strategy2}
for $\bar c'$ as in the footnote there. 
By this choice, we can set $\bar c'=-c-z_i$, instead of 
 $\bar c'=-c-z_C-z_i$, so that 
 $\bar C'\Ls A+Z\Rs C$ is allowed. 
This can reduce $\bar m$, and 
 we can construct a model that suppresses 
 the FCNC process to the same level as in the models introduced 
 in \S\ref{strategy2} and is able to be embedded 
 into $SU(3)_H$ model. (See Table \ref{SU2ForSU3-2})
Actually, the parameters $\delta_{\bf 10}$ and $\delta_{\bf \bar 5}$
are the same expression as in the Eqs. (\ref{e6su2delta}).
\begin{table}
\caption{
An example of the charge assignments of $SU(2)_H$ models that 
 can be embedded into $SU(3)_H$ models : 
Signs denote the $Z_2$ symmetry
that play the same role as the $Z_2$ symmetry 
 introduced in Table \ref{HiggsContentE6}. 
We impose an additional $Z_2$ symmetry 
 and introduce a singlet field $Z_C$. 
The FCNC processes are suppressed to the same level as in models 
 in Table \ref{ContentStrategy12}.
This charge assignment yields $r=1$ and 
 $l=-5/2$.
Odd quarter integer charges of the matter fields ($\Psi_3$, $\Psi_a$) 
 guarantees that the $R$-parity is automatically conserved.
}
\label{SU2ForSU3-2}
\[
\begin{array}{|c|c|c|} 
\hline
                  &   \mbox{non-vanishing VEV}  
                  &   \mbox{vanishing VEV} \\
\hline 
{\bf 78}          &   A(a=-1;-)        & A'(a'=5;-)  \\
{\bf 27}          &   \Phi(\phi=-7/2;+),\,  C(c=-5/2;+) 
                  &   C'(c'=8;-)  \\
                 &&   \Psi_3 (\psi_3=7/4;+),\,\Psi_a (\psi=19/4;+)\\
{\bf \cc{27}}     &   \bar C_a(\bar c=1;+) 
                  &   \bar C'(\bar c'=4;-) \\
{\bf 1}           &   \bar F_a (\bar f=-2;+),\,F_a(f=-1/2;+) &\\
                  &   \Theta (\theta=-1/2;+),\, Z_i(z_i=-3/2;-) &  \\
                  &   Z_C(z_C=-3/2;+) &  \\
\hline
\end{array}
\]
\end{table}

\subsubsection{$SU(3)_H$ models}

Now, we treat $SU(3)_H$ models. 
The Higgs content is summarized in Table \ref{HiggsContentE6SU3}%
\footnote{
Note that the $SU(3)_H$ horizontal symmetry in this model 
 is anomaly free.
}.
\begin{table}[th]
\caption{
The Higgs content of $E_6\times SU(3)_H\times U(1)_A$ models 
 expect for singlets:
Here $SU(3)_H$ triplets and anti-triplets are denoted 
 by the lower and upper index $\alpha$, respectively. 
All the non-vanishing VEVs are shown, and their magnitudes 
 are formally written by introducing parameters $\Delta\phi$ etc..
One or more discreate symmetries are introduced according to need.
}
\label{HiggsContentE6SU3}
\[
\begin{array}{|c|c|c|} 
\hline
           &   \mbox{non-vanishing VEV}  & \mbox{vanishing VEV} \\
\hline 
  {\bf 78} &   A\Ls \VEV{{\bf45}_A}\sim\lambda^{-a}\Rs  & A'      \\
  {\bf 27} &   \Phi\Ls \VEV{{\bf1}_\Phi}
                   \sim\lambda^{-(\phi-\Delta\phi)}\Rs,\ 
               C\Ls \VEV{{\bf16}_C}\sim\lambda^{-(c-\Delta c)}\Rs 
           &  C'  \\
  {\bf\cc{27}} & \bar C^\alpha\Ls 
                    \VEV{{\bf\cc{16}}_{\bar C_1}}
               \sim \lambda^{-(\bar c+\Delta c-\Delta f-\Delta e/2)},\ 
                    \VEV{{\bf\cc{1}}_{\bar C_2}}
               \sim \lambda^{-(\bar c+\Delta\phi+\Delta f-\Delta e/2)} \Rs 
               & \\
  {\bf1} & F_\alpha\Ls \VEV{F_2}\sim\lambda^{-(f-\Delta f+\Delta e/2)}\Rs,\ 
           \bar F^\alpha \Ls \VEV{\bar F_2}
                    \sim\lambda^{-(\bar f+\Delta f-\Delta e/2)}\Rs& \\
         & E_\alpha\Ls \VEV{E_3}\sim\lambda^{-(e-\Delta e)}\Rs,\ 
           \bar E^\alpha \Ls \VEV{\bar E_3}
                    \sim\lambda^{-(\bar e+\Delta e)}\Rs& \\
\hline
\end{array}
\]
\end{table}
Each component of a triplet $\Psi_\alpha$ and 
 an anti-triplet $\bar C^\alpha$
 can be picked up as 
\beqn
 (\Psi_1, \Psi_2, \Psi_3) &\sim& (\Psi EF, \Psi\bar F, \Psi E) 
\label{triplet}\\
 (\bar C_1, \bar C_2, \bar C_3) &\sim& 
    (\bar C\bar E\bar F, \bar CF, \bar CE)
\label{antitriplet},
\eeqn
and the effective charge of each element is given as
\beqn
 \s\psi &=& (\psi+\Delta f+\Delta e/2, \psi-\Delta f+\Delta e/2, 
           \psi-\Delta e) \\
 \s{\bar c} &=& (\bar c-\Delta f-\Delta e/2, \bar c+\Delta f-\Delta e/2, 
                 \bar c+\Delta e) .
\eeqn
This means that, providing $e=-\bar e=\Delta e$ and 
 integrating out $E$ and $\bar E$, 
 we get a $SU(2)_H$ model where 
 $(\psi, \psi_3, \bar c, \bar c', \bar f, f)$ are given as 
 $(\psi+e/2, \psi_3-e, \bar c-e/2, \bar c'+e, \bar f-e/2, f+e/2)$ 
 in terms of the charges in the $SU(3)_H$ model%
\footnote{
As for the $Z_2$-parities, we can find those of each component 
 from Eqs.(\ref{triplet}) and (\ref{antitriplet}). 
In addition, for example,  $\bar C\bar E\bar C\Phi$ and 
 $\bar C\bar E\bar FZ_C$ (whose charges are usually smaller 
 than that of $\bar C\bar E\bar F$) 
 pick up $C_1$ component with 
 another parity assignments 
 (they may be same as the previous one). 
}. 
Conversely, we can construct an $SU(3)_H$ model with $e=-e=2$ 
 as shown in Table \ref{ContentSU3} from a $SU(2)_H$ model in 
 Table \ref{SU2ForSU3-2}. 
\begin{table}[ht]
\caption{
An example of the charge assignments of $SU(3)_H$ models : 
Signs denote the $Z_2$ symmetry 
 that play the same role as the $Z_2$ symmetry 
 introduced in Table \ref{HiggsContentE6}. 
We impose an additional $Z_2$ symmetry 
 and introduce a singlet field $Z_C$. 
The FCNC processes are suppressed to the same level as in models 
 in Table \ref{ContentStrategy12}.
This charge assignment yields $r=1$ and 
 $l=-5/2$.
Odd quarter integer charge of the matter field ($\Psi_\alpha$) 
 guarantees that the $R$-parity is automatically conserved.
}
\label{ContentSU3}
\[
\begin{array}{|c|c|c|} 
\hline
                  &   \mbox{non-vanishing VEV}  
                  &   \mbox{vanishing VEV} \\
\hline 
{\bf 78}          &   A(a=-1;-)        & A'(a'=5;-)  \\
{\bf 27}          &   \Phi(\phi=-7/2;+),\,  C(c=-5/2;+) 
                  &   C'(c'=8;-)  \\
                 &&   \Psi_\alpha (\psi=15/4;-) \\
{\bf \cc{27}}     &   \bar C^\alpha(\bar c=2;+) 
                  & \\
{\bf 1}           &   F_\alpha (f=-3/2;+),\,
                      \bar F^\alpha(\bar f=-1;-) &\\
                  &   E_\alpha (e=2;-),\,
                      \bar E^\alpha(\bar e=-2;-) &\\
                  &   \Theta (\theta=-1/2;+),\, Z_i(z_i=-3/2;-) &  \\
                  &   Z_C(z_C=-3/2;+) &  \\
\hline
\end{array}
\]
\end{table}
Here, parity assignment of the additional $Z_2$ symmetry 
 for (anti)triplet fields 
 $(\Psi,\bar C,F,\bar F,E,\bar E)$ is 
 $(-,+,+,-,-,-)$, 
 so that $\bar C_a\,(a=1,2)$ and $\Psi_\alpha\,(\alpha=1,2,3)$ 
 have even parity while $\bar C_3$ has odd parity, 
 and the others have the same parity as in the $SU(2)_H$ model. 
This parity plays essentially the same role 
 as that in the $SU(2)_H$ model in Table \ref{SU2ForSU3-2}.

\section{Summary and Discussion} 
In this paper, 
we have investigated $SO(10)$ or $E_6$ SUSY-GUTs with 
 an anomalous $U(1)$ symmetry and an $SU(2)_H$ or $SU(3)_H$ 
 horizontal symmetry, where some of GUT-breaking Higgs belong to 
 non-trivial representations of the horizontal symmetry. 
We have found it possible to unify the Higgs sectors for the GUT symmetry and
the horizontal symmetry. It is interesting that for $SU(3)_H$ models, $SU(3)_H$
gauge anomaly is cancelled between the triplet matter $\Psi_a$ and the 
anti-triplet Higgs $\bar C^a$.

Unfortunately, the unification of the Higgs sectors of the GUT symmetry and
the horizontal symmetry results in too large FCNC processes. 
This is because in the scenario of the GUT with the anomalous $U(1)$ symmetry
the cutoff scale, $\Lambda$, must be around the usual GUT scale, 
$2\times 10^{16}$ GeV, to realize the natural gauge coupling unification 
\cite{NGCU} and
the true GUT scale in our scenario is difficult to be smaller than 
$\lambda^2\Lambda$, which
is the sufficient value for suppressing the FCNC processes. 
At present, we do not know the meaning of this fact.
This fact may mean that another mechanism is required to realize the 
universality of sfermion masses, or that the fields in the Higgs sector of
the GUT symmetry do not have non-trivial quantum numbers under the horizontal 
symmetry, or that the anomalous $U(1)_A$ is given up although the GUT models
with the anomalous $U(1)_A$ solve the doublet-triplet splitting problem and
 realize the natural gauge coupling unification with generic interactions.
However, we hope that the arguments in this paper give a hint 
in finding the real grand unified theory, which we expect to be just around 
the corner.

\section*{Acknowledgement}
This work is supported by a Grant-in-Aid for 
 the 21st Century COE''Center for Diversity 
 and Universality in Physics''.
N.M. is supported in part by Grants-in-Aid for Scientific 
Research from the Ministry of Education, Culture, Sports, Science 
and Technology of Japan.

\appendix

\section{Effects of deviation from effective charge}
\label{diff}

In this appendix, we analyse the model defined 
 in Table \ref{ContentStrategy3}, in which the VEV relations (\ref{VEV})
 are not satisfied, that is, the effective charges are not well defined.

\subsection{${\bf5}$-${\bf{\bar5}}$ mass matrix of matters}
\label{55barmass}

First, we consider the mass matrix of ${\bf5}$-${\bf{\bar5}}$ 
 components of $\Psi_a$ and $\Psi_3$.
Because of the decomposition of the fundamental representation of 
 $E_6$ (\ref{27}), 
 this matrix becomes $3\times6$.
The mass terms of $({\bf{10}},{\bf5})$-$({\bf{10}},{\bf{\bar5}})$  
 are given from $\Psi\Psi\VEV\Phi$ and those of 
 $({\bf{10}},{\bf5})$-$({\bf{16}},{\bf{\bar5}})$ are given from
 $\Psi\Psi\VEV C$. 
For the charge assignment in Table \ref{ContentStrategy3}, 
 this matrix is given in terms of the exponent as
\begin{equation}
  \bordermatrix{
      & \Psi_1({\bf 10,\bar 5}) & \Psi_2({\bf 10,\bar 5})
        & \Psi_3({\bf 10,\bar5})
      & \Psi_1({\bf 16, \bar 5}) & \Psi_2({\bf 16, \bar5})
        & \Psi_3({\bf 16,\bar 5})\cr 
    \Psi_1({\bf 10,5}) 
      & 6+\delta & 5
        & 3+\delta 
      & 6+\delta+r &  5+r
        & 3+\delta+r  \cr 
    \Psi_2({\bf 10,5})
      & 5    &  4   & 2
      & 5+r  &  4+r & 2+r \cr  
    \Psi_3({\bf 10,5}) 
      & 3+\delta & 2
        & 0
      & 3+\delta+r & 2+r
        & \mbox{-} 
   } -k,
\label{55barmassmatrix}
\end{equation}
where ``-'' means that the corresponding element of the $3\times 6$ mass matrix
is forbidden by the SUSY-zero mechanism.
From this matrix, we can find the three ${\bf{\bar5}}$ 
 that remain massless are given as
\beqn
 {\bf\bar5_1}&=&\Psi_1({\bf 16, \bar 5})+
                \lambda^3\Psi_3({\bf 16, \bar 5})+
                \lambda^{3+\delta+r}\Psi_3({\bf 10, \bar 5})+
                \lambda^{1+\delta+r}\Psi_2({\bf 10, \bar 5}),  \\
 {\bf\bar5_2}&=&\Psi_1({\bf 10, \bar 5})+
                \lambda^{3-r}\Psi_3({\bf 16, \bar 5})+
                \lambda^{3+\delta}\Psi_3({\bf 10, \bar 5})+
                \lambda^{1+\delta}\Psi_2({\bf 10, \bar 5}),  \\
 {\bf\bar5_3}&=&\Psi_2({\bf 16, \bar 5})+
                \lambda^2\Psi_3({\bf 16, \bar 5})+
                \lambda^{2+r}\Psi_3({\bf 10, \bar 5})+
                \lambda^{r}\Psi_2({\bf 10, \bar 5}). 
\eeqn

In order to obtain the Yukawa couplings of quarks and leptons, we must know
the origin of the MSSM Higgs $H_u$ and $H_d$.
Here we write just the results:
\bequ
 H_u\sim {{\bf 10}_\Phi}+\lambda^\half {{\bf 10}_C},\quad 
 H_d\sim {{\bf{16}}_C}+\lambda^{c-\phi+r}{\bf10}_\Phi
        +\lambda^{r}{\bf10}_C, 
\eequ
which will be shown in Appendix \ref{Higgs-mass}.
Then we obtain the following Yukawa matrix and mixing matrix for 
 the ${\bf\bar5}$ sector as
\bequ
 y_{\bf\bar5}=\Ls
  \begin{array}{ccc}
    \lambda^{6+\delta}     & \lambda^5     & \lambda^3 \\
    \lambda^{6+\delta-r} & \lambda^{5-r} & \lambda^{3-r} \\
    \lambda^5     & \lambda^4     & \lambda^2
  \end{array}
  \Rs\lambda^{r-\half},\quad
 V_{\bf\bar5}=\Ls
  \begin{array}{ccc}
    1         & \lambda^r     & \lambda\\
    \lambda^r & 1             & \lambda^{1-r} \\
    \lambda   & \lambda^{1-r} & 1
  \end{array}
  \Rs.
\eequ

Note that if $\Psi_3\Psi C\bar F$ is not allowed, 
 the (2,6) and (3,5) elements of the matrix (\ref{55barmassmatrix}) 
 become ``-''. 
Then the light three $\bf{\bar5}$ modes become 
\beqn
 {\bf\bar5_1}&=&\Psi_1({\bf 16, \bar 5})+
                \lambda^{3-\delta}\Psi_3({\bf 16, \bar 5})+
                \lambda^{3+r}\Psi_3({\bf 10, \bar 5})+
                \lambda^{1+r}\Psi_2({\bf 10, \bar 5}),  \\
 {\bf\bar5_2}&=&\Psi_1({\bf 10, \bar 5})+
                \lambda^{3-\delta-r}\Psi_3({\bf 16, \bar 5})+
                \lambda^{3}\Psi_3({\bf 10, \bar 5})+
                \lambda^{1}\Psi_2({\bf 10, \bar 5}),  \\
 {\bf\bar5_3}&=&\Psi_2({\bf 16, \bar 5})+
                \lambda^{2-\delta}\Psi_3({\bf 16, \bar 5})+
                \lambda^{2+r}\Psi_3({\bf 10, \bar 5})+
                \lambda^{r}\Psi_2({\bf 10, \bar 5}). 
\eeqn
which leads the Yukawa matrix to 
\bequ
 y_{\bf\bar5}=\Ls
  \begin{array}{ccc}
    \lambda^{6+\delta}     & \lambda^5     & \lambda^3 \\
    \lambda^{6+\delta-r} & \lambda^{5-r} & \lambda^{3-r} \\
    \lambda^{5+\delta}     & \lambda^4     & \lambda^2
  \end{array}
  \Rs\lambda^{r-\half-\delta}
\eequ
yielding the ratio between the 1st and 3rd eigenvalues as 
 $m_{\bf{\bar5}_1}/m_{\bf{\bar5}_3}\sim\lambda^{4+\delta}$ 
 which is a bit too small because $\delta=\frac{4}{3}$.
Thus, the term $\Psi_3\Psi C\bar F$ is required.

\subsection{Neutrino mass}

Next, we consider the neutrino mass matrix.
Their Yukawa couplings are give by $3\times6$ matrix because 
 there are two right-handed neutrinos in each ${\bf{27}}$.
They are given from 
 $\Psi({\bf{16}},{\bf{\bar5}})\Psi({\bf{16}},{\bf1})\Phi({\bf{10}},{\bf5})$ 
 and 
 $\Psi({\bf{10}},{\bf{\bar5}})\Psi({\bf1},{\bf1})\Phi({\bf{10}},{\bf5})$. 
The Yukawa matrix is given as
\bequ
  \bordermatrix{
      &\Psi_1({\bf 16,1})&\Psi_2({\bf 16,1}) &\Psi_3({\bf16,1})
      &\Psi_1({\bf 1,1})&\Psi_2({\bf 1,1})&\Psi_3({\bf1,1})   \cr
    {\bf \cc 5}_1 
      & \lambda^{6+\delta}   & \lambda^5     & \lambda^3
      & \lambda^{6+\delta+r} & \lambda^{5+\delta+r} 
                             & \lambda^{3+\delta+r}    \cr
    {\bf \cc 5}_2  
      & \lambda^{6+\delta-r} & \lambda^{5-r} & \lambda^{3-r}
      & \lambda^{6+\delta}   & \lambda^{5+\delta}  
                             & \lambda^{3+\delta}      \cr
    {\bf \cc 5}_3  
      & \lambda^5            & \lambda^4     & \lambda^2
      & \lambda^{5+r}        & \lambda^{4+r}       
                             & \lambda^{2+r}           \cr
  }.
\label{nuYukawa}
\eequ
Note that this is written as 
\bequ
  \Ls\begin{array}{cc}
       \lambda^{2}   & \lambda^{2+r} \cr
  \end{array}\Rs
  \otimes
  \Ls\begin{array}{ccc}
       \lambda^{4+\delta}   & \lambda^3     & \lambda^1  \cr
       \lambda^{4+\delta-r} & \lambda^{3-r} & \lambda^{1-r}\cr
       \lambda^3            & \lambda^2     & 1  \cr
  \end{array}\Rs,
\label{nuYukawaAppro}
\eequ
except for the (1,5), (1,6), (2,5) and (2,6) elements, which are 
 smaller than the corresponding elements of 
 the expression (\ref{nuYukawaAppro}).

The Majorana mass matrix of the right-handed neutrinos, $M_R$, is 
 a $6\times6$ matrix and their elements are given from 
 $\Psi\Psi\VEV{\bar C\bar C}$. 
This matrix is written in terms of the exponent as
{\footnotesize{
\beqn
  &\bordermatrix{
      &\Psi_1({\bf 16,1})&\Psi_2({\bf 16,1}) &\Psi_3({\bf16,1})
      &\Psi_1({\bf 1,1})&\Psi_2({\bf 1,1})&\Psi_3({\bf1,1})   \cr
    \Psi_1({\bf 16,1})
      & 6-2r-\delta  &  5-2r-2\delta  &  3-2r-\delta
      & 6- r-\delta  &  5- r-2\delta  &  3- r-\delta    \cr
    \Psi_2({\bf 16,1})
      & 5-2r-2\delta  &  4-2r         &  2-2r
      & 5- r-2\delta  &  4-r-3\delta  &  2- r-2\delta      \cr
    \Psi_3({\bf 16,1})
      & 3-2r-\delta  &  2-2r          & \mbox{-}
      & 3- r-\delta  &  2- r-2\delta  &   -r-\delta           \cr
    \Psi_1({\bf 1,1})
      & 6-r-\delta  &  5-r-2\delta  &  3-r-\delta
      & 6           &  5  -2\delta  &  3           \cr
    \Psi_2({\bf 1,1})
      & 5-r-2\delta  &  4-r-3\delta  &  2-r-2\delta
      & 5  -2\delta  &  4  -3\delta  &  2  -2\delta      \cr
    \Psi_3({\bf 1,1})
      & 3-r-\delta  &  2-r-2\delta  & -r-\delta  
      & 3           &  2  -2\delta  &   -\delta           \cr
  }&\nn\\&+3.5-2\Delta\phi\hsp{10}&, 
\label{Majorana}
\eeqn
}}
and can be expressed as
\bequ
  \lambda^{3.5-2\Delta\phi-\delta}
  \Ls\begin{array}{cc}
       \lambda^{-2r}   & \lambda^{-r} \cr
       \lambda^{- r}   & 1            \cr
  \end{array}\Rs
  \otimes
  \Ls\begin{array}{ccc}
       \lambda^{6}   & \lambda^{5-\delta}     & \lambda^3  \cr
       \lambda^{5-\delta} & \lambda^{4-2\delta} & \lambda^{2-\delta}\cr
       \lambda^3            & \lambda^{2-\delta}     & 1  \cr
  \end{array}\Rs,
\label{MajoranaAppro}
\eequ
except for the (2,2), (2,3), (3,2), (3,3), (4,4), (4,6) and (6,4) 
 elements, which are 
 smaller than the corresponding elements of 
 the expression (\ref{MajoranaAppro}).
These small elements have only sub-leading contributions to 
 the inverse matrix $M_R^{-1}$.
Also, the small elements in the expression (\ref{nuYukawa}) 
 have only sub-leading contributions to the light neutrino mass.
Thus, the correct order of magnitudes of the light neutrino mass
 are obtained by a calculation using the expressions 
 (\ref{nuYukawaAppro}) and (\ref{MajoranaAppro}) instead of 
 (\ref{nuYukawa}) and (\ref{Majorana}), which leads to 
\beqn
&\lambda^{-3.5+2\Delta\phi+\delta}\VEV{H_u}^2 
  \Ls\begin{array}{cc}
       \lambda^{2}   & \lambda^{2+r} \cr
  \end{array}\Rs
  \Ls\begin{array}{cc}
       \lambda^{2r}   & \lambda^{r} \cr
       \lambda^{ r}   & 1            \cr
  \end{array}\Rs
  \Ls\begin{array}{c}
       \lambda^{2}  \cr \lambda^{2+r} \cr
  \end{array}\Rs 
  \otimes
&\nn\\
&  \Ls\begin{array}{ccc}
       \lambda^{4+\delta}   & \lambda^3     & \lambda^1  \cr
       \lambda^{4+\delta-r} & \lambda^{3-r} & \lambda^{1-r}\cr
       \lambda^3            & \lambda^2     & 1  \cr
  \end{array}\Rs
  \Ls\begin{array}{ccc}
       \lambda^{-6}   & \lambda^{-5+\delta}     & \lambda^{-3}  \cr
       \lambda^{-5+\delta} & \lambda^{-4+2\delta} & \lambda^{-2+\delta}\cr
       \lambda^{-3}            & \lambda^{-2+\delta}     & 1  \cr
  \end{array}\Rs
  \Ls\begin{array}{ccc}
       \lambda^{4+\delta} & \lambda^{4+\delta-r} & \lambda^3  \cr
       \lambda^3          & \lambda^{3-r}        & \lambda^2\cr
       \lambda^1          & \lambda^{1-r}        & 1  \cr
  \end{array}\Rs
&\label{nuMass}\\
&
=
\lambda^{-3.5+2\Delta\phi+\delta}
\lambda^{4+2r}
  \Ls\begin{array}{ccc}
       \lambda^{2}   & \lambda^{2-r} & \lambda^{1}  \cr
       \lambda^{2-r} & \lambda^{2-2r}  & \lambda^{1-r}\cr
       \lambda^{1}   & \lambda^{1-r}   & 1  \cr
  \end{array}\Rs \VEV{H_u}^2 .
&\label{nuMass2}\eeqn
In this case, the parameter $l$ is given as 
\bequ
  -(l+5)=-3.5+2\Delta\phi+\delta+4+2r.
\eequ
This is different from the previous expression for $l$ (\ref{2nd}) by $\delta$.
Note that the determinant of the matrix (\ref{nuMass}) is smaller than
the naively expected determinant of the matrix (\ref{nuMass2}) 
 by a factor $\lambda^{2\delta}$.
This means a cancellation must occur 
 in caluculating the eigenvalues of the matirx (\ref{nuMass2}), 
 and the mass of the lightest neutrino become,
\bequ
 \frac{m_{\nu_1}}{m_{\nu_3}}\sim \lambda^{2+2\delta},
\eequ
which is smaller than the naively expected value from the
matrix (\ref{nuMass2}),  
$\frac{m_{\nu_1}}{m_{\nu_3}}\sim \lambda^{2}$.

\subsection{Mass matrix of GUT-breaking Higgs}
\label{Higgs-mass}

Finally we examine mass matrices of Higgs that break the $E_6$ gauge 
symmetry. Because the VEV relations (\ref{VEV}) are not satisfied in 
this model,
the success of the gauge coupling unification may be spoiled. Actually,
the coefficients of some effective interactions are dependent on the 
original interactions given at the cutoff scale as discusses in 
\S\ref{strategy}, that is, the effective charges cannot be defined well.
Moreover, if the total charge of a interaction is large enough for 
the singlet operator,
$\bar C\bar C\Phi C$, to couple with the interaction, 
the coefficient for the interaction
is enhanced by a factor 
$\lambda^{2\bar c+\phi+c}\VEV{\bar C\bar C\Phi C}\sim \epsilon^{-1}$.
Such effects, in principle, disturb the gauge coupling unification,
which are guaranteed if the effective charges can be well-defined\cite{NGCU}. 
Let us illustrate these by calculating the mass matrix of the 
 ${\bf5}$-${\bf{\bar5}}$ components of Higgs explicitly as an example.
It is given as following 11$\times$11 matrix.
\begin{equation}
  \bordermatrix{
    \bar I\backslash I 
      & {\bf10}_C & {\bf10}_\Phi & {\bf16}_C 
      & {\bf16}_\Phi & {\bf16}_A 
            & {\bf16}_{A'} & {\bf10}_{\bar C_2} & {\bf10}_{\bar C_1} 
            & {\bf16}_{C'} 
      & {\bf10}_{C'} & {\bf10}_{\bar C'}  \cr
    {\bf10}_C
      & 0 & 0 & 0 & 0 
      & 0 & 0 & 0 & 0 
      & 0 & 0 & \lambda^{\frac{11}{2}}  \cr
    {\bf10}_\Phi
      & 0 & 0 & 0 & 0 
      & 0 & 0 & 0 & 0 
      & 0 & \alpha_I \lambda^{\frac{4}{3}} & \lambda^{6}  \cr
    \cr
    {\bf{\cc{16}}}_{\bar C_2}
      & 0 & 0 & 0 & 0 
      & 0 & 0 & 0 & 0 
      & \frac{1}{\epsilon}\lambda^{7} 
            & 0
            & \frac{1}{\epsilon^2}\lambda^{\frac{43}{4}}  \cr
    {\bf10}_{\bar C_2}
      & 0 & 0 & 0 & 0 
      & 0 & 0 & 0 & \lambda^{4} 
      & \frac{1}{\epsilon}\lambda^{\frac{95}{12}} 
            & \frac{1}{\epsilon}\lambda^{7} 
            & \frac{1}{\epsilon^2}\lambda^{\frac{35}{3}}  \cr
    {\bf{\cc{16}}}_{\bar C_1}
      & 0 & 0 & 0 & 0 
      & 0 & \lambda^{\frac{13}{6}} & 0 & \lambda^{\frac{65}{12}} 
      & \frac{1}{\epsilon}\lambda^{\frac{28}{3}} 
            & \frac{1}{\epsilon}\lambda^{\frac{101}{12}} 
            & \frac{1}{\epsilon^2}\lambda^{\frac{157}{12}}  \cr
    {\bf{\cc{16}}}_A
      & 0 & 0 & 0 & 0 
      & 0 & \lambda^{\frac{5}{2}}& 0 & 0 
      & \frac{1}{\epsilon}\lambda^{\frac{29}{3}} 
          & \frac{1}{\epsilon}\lambda^{\frac{35}{4}}
          & \frac{1}{\epsilon^2}\lambda^{\frac{161}{12}}  \cr
    {\bf10}_{\bar C_1}
      & 0 & 0 & 0 & 0 
      & 0 & 0 & \lambda^{4} & \lambda^{\frac{19}{3}} 
      & \frac{1}{\epsilon}\lambda^{\frac{41}{4}} 
            & \frac{1}{\epsilon}\lambda^{\frac{28}{3}} 
            & \frac{1}{\epsilon^2}\lambda^{14}  \cr
    {\bf{\cc{16}}}_{A'}
      & 0 & 0 & 0 & \lambda^{\frac{13}{6}} 
      & \lambda^{\frac{5}{2}} & \lambda^{6} & 0 & 0 
      & \frac{1}{\epsilon}\lambda^{\frac{79}{6}} 
            & \frac{1}{\epsilon}\lambda^{\frac{49}{4}} 
      & \frac{1}{\epsilon^2}\lambda^{\frac{203}{12}}  \cr
    \cr
    {\bf10}_{C'}
      & 0 & \alpha_I \lambda^{\frac{4}{3}} & 0 & 0 
      & 0 & \lambda^{\frac{73}{12}} 
            & \frac{1}{\epsilon}\lambda^{7}
            & \frac{1}{\epsilon}\lambda^{\frac{28}{3}} 
      & \frac{1}{\epsilon}\lambda^{\frac{53}{4}} 
            & \frac{1}{\epsilon}\lambda^{\frac{37}{3}} 
            & \frac{1}{\epsilon^{2}}\lambda^{17}  \cr
    {\bf{\cc{16}}}_{\bar C'}
      & 0 & \lambda^{\frac{61}{12}} & \lambda^{\frac{11}{2}} 
            & \lambda^{6} 
      & \lambda^{\frac{19}{3}} & \lambda^{\frac{59}{6}} 
            & \frac{1}{\epsilon^{2}}\lambda^{\frac{43}{4}}
            & \frac{1}{\epsilon^{2}}\lambda^{\frac{157}{12}}
      & \frac{1}{\epsilon^{2}}\lambda^{17} 
            & \frac{1}{\epsilon^{2}}\lambda^{\frac{193}{12}} 
            & \frac{1}{\epsilon^{3}}\lambda^{\frac{83}{4}}\cr
    {\bf10}_{\bar C'}
      & \lambda^{11/2} & \lambda^{6} & \lambda^{\frac{77}{12}} 
            & \lambda^{\frac{83}{12}}
      & \lambda^{\frac{29}{4}} &\lambda^{\frac{43}{4}}
            & \frac{1}{\epsilon^{2}}\lambda^{\frac{35}{3}}
            & \frac{1}{\epsilon^{2}}\lambda^{14}
      & \frac{1}{\epsilon^{2}}\lambda^{\frac{215}{12}} 
            & \frac{1}{\epsilon^{2}}\lambda^{17} 
            & \frac{1}{\epsilon^{3}}\lambda^{\frac{65}{3}}  \cr
  }
\label{E6DL}
\end{equation}
Here, $\alpha_I=0$ for the doublet component, which yields 
 additional massless mode only for doublet written as 
\bequ
 H_u\sim {\bf10}_\Phi + \lambda^{\phi-c}{\bf10}_C,\qquad 
 H_d\sim {{\bf{16}}_C}+\lambda^{c-\phi+r}{\bf10}_\Phi
        +\lambda^{r}{\bf10}_C, 
\eequ
which are identified with the MSSM doublet Higgs.

In principle, because the VEV realtions are not satisfied,
the gauge coupling unification is not guranteed in this model.
Of course, the discrepancy is parametrized by $\epsilon$,
because if $\epsilon\sim \order1$, the natural gauge coupling
unification is realized as discussed in Refs. \cite{NGCU}.
In fact, the mass spectrum, which is important in estimating the
discrepancy, are given as 
 $m_T=\lambda^{\frac{11}{2}},\lambda^{\frac{4}{3}},
    {\frac{1}{\epsilon}}\lambda^{7},\lambda^4,
    \lambda^{\frac{13}{6}},0,\lambda^4,
    \lambda^{\frac{13}{6}},\lambda^{\frac{4}{3}},
    \lambda^{\frac{11}{2}},\lambda^{\frac{11}{2}}$ 
 for the triplet components and 
 $m_D=\lambda^{\frac{11}{2}},0,
    {\frac{1}{\epsilon}}\lambda^{7},\lambda^4,
    \lambda^{\frac{13}{6}},0,\lambda^4,
    \lambda^{\frac{13}{6}},
    {\frac{1}{\epsilon}}\lambda^{\frac{37}{3}},
    \lambda^{\frac{61}{12}},\lambda^{\frac{11}{2}}$ 
 for the doublet components.
Note that $1/\epsilon$ appear in relatively 
 small elements. This is mainly because the enhancement factor,  
$\lambda^{2\bar c+\phi+c}\VEV{\bar C\bar C\Phi C}\sim \epsilon^{-1}$,
can appear in the terms with sufficiently large anomalous $U(1)_A$ charges.
The ratio $\det m_D/\det m_T$, which is important 
 in estimating the discrepancy, is enhanced by a factor $\epsilon^{-1}$
 in this model.
This enhancement improves the gauge coupling unification because relatively
light colored Higgs mass (and therefore, heavy doublet Higgs mass) is prefered.
This feaure that the ratio is enhanced and the gauge coupling unification
is improved, is comparatively general in this scenario.
The reasons are as follows.
The typical mass matrix for a $SU(5)$ irreducible representation fields,
$X$ and $X'$ ($\bar X$ and $\bar X'$ are conjugate representation fields)
is given as
\begin{equation}
\left(\begin{array}{cc}
        0 & \lambda^{x+\bar x'}\alpha \\
        \lambda^{\bar x+x'}\alpha & \lambda^{\bar x'+x'} 
      \end{array}\right),
\end{equation}
where the effective charges $x$ and $\bar x$ are negative and $x'$ and 
$\bar x'$ are positive. The $\order1$ paramter $\alpha$ is vanishing for
one of the component fields for $X$, $\bar X$, $X'$, and $\bar X'$.
( For ${\bf \bar 5}$ and ${\bf 5}$ fields, $\alpha=0$ for the doublet
Higgs because of the doublet-triplet splitting.  In our scenario, 
such vanishing parameters like $\alpha$ also appear 
for the Nambu-Goldstone (NG) modes in breaking $E_6\rightarrow G_{\rm SM}$.)
Bacause $x$ and $\bar x$ are negative and $x'$ and $\bar x'$ are positive, 
the (1,1) element is vanishing by
the SUSY zero mechanism and the non-diagonal elements are larger than the
other diagonal (2,2) element.
Therefore, when $\alpha\neq 0$, the determinant of the mass matrix is given by 
the product of the non-diagonal elements, $\lambda^{x+\bar x+x'+\bar x'}$.
However, when $\alpha=0$, the mass eigenvalues become 0 and  
$\lambda^{x'+\bar x'}$, which are the smallest non-vanishing eigenvalue
among the mass spectrum of $X$, $X'$, $\bar X$, and $\bar X'$. 
As discussed in the above, it is plausible that the factor $\epsilon^{-1}$ 
appears in the (2,2) element, which is enhanced if $\epsilon\ll1$,
that is, the eigenvalue for the component field
with the vanishing $\alpha$ is enhanced.
For the mass matrix of ${\bf\bar 5}$ and ${\bf 5}$ fields, if doublet-triplet
splitting is realized by a certain mechanism, the mass eigenvalues of doublet
components tend to be enhanced when $\epsilon \ll1$. (The effect of the NG 
modes can be neglected in many cases because the content of the NG modes 
respects $SU(5)$ symmetry.)
For the mass matrices of the other representation fields, ${\bf 10}$
and ${\bf 24}$ of $SU(5)$, we can roughly discuss the effect of
$\epsilon\ll1$. 
For the mass matrix of ${\bf 10}$ and ${\bf \overline{10}}$ fields, the content
of the NG modes respects the $SU(5)$. 
Therefore, the effect of $\epsilon\ll1$ can be small in many cases, althogh
the effect depends on the concrete models. 
(In the model in Table \ref{ContentStrategy3}, the effect can be neglected.) 
For the mass matrix of ${\bf 24}$ fields, the NG modes are
$({\bf 3,2})_{-\frac{5}{6}}$ and $({\bf \bar 3,2})_{\frac{5}{6}}$ under 
the SM gauge group. Therefore, these modes become heavier when $\epsilon\ll1$.
In summary, the effect of the deviation from the effective charge, 
 $\epsilon$, on the gauge coupling unification appear in mass 
 spectrums of the doublet and $({\bf 3,2})_{-\frac{5}{6}}$ pairs 
 by enhancement factors, $\epsilon^{-1}$, in this model. 
The enhancement factor in the spectrum of the doublet reduces 
 the disagreement 
 of the gauge couplings caused by the large effective mass 
 of the colored Higgs, $m_\eff\gg\Lambda$.

\end{document}